\documentclass[iop]{emulateapj}
\slugcomment{{\sc Accepted to ApJ:} May 17, 2011} 
\usepackage{amssymb,amsmath}
\newcommand\nion[2]{#1\,\lowercase{{\sc #2}}}
\newcommand\wave[1]{\mbox{$\lambda$#1\,\AA}}
\def\kmsec{${\rm km~s^ {-1}}$}
\def\BmV0{\mbox{$(B-V)^{\rm 0}$}}
\def\VmK0{\mbox{$(V-K)^{\rm 0}$}}
\def\MV0{\mbox{$M_{\rm V}^{\rm 0}$}}

\def\msun{${\rm M}_{\odot}$}

\def\simgt{\lower.5ex\hbox{$\; \buildrel > \over \sim \;$}}
\def\simlt{\lower.5ex\hbox{$\; \buildrel < \over \sim \;$}}
\def\teff{T_{\rm eff}}
\def\tini{T_{\rm ini}}
\def\tphot{T_{\rm phot}}
\def\logg{\log~g}
\def\loggi{\log~g_{\rm ini}}
\def\loggb{\log~g_{\rm bol}}
\def\loggp{\log~g_{\rm phot}}
\def\fehi{\rm[Fe/H]_{ini}}
\def\feh{\rm[Fe/H]}
\def\mh{\rm[M/H]}
\def\afe{\rm[\alpha/Fe]}

\def\ks{K_{\rm S}}
\def\mks{M_{K_{\rm S}}}

\def\aks{A_{K_{\rm S}}}
\def\rapo{r_{\rm apo}}
\def\rper{r_{\rm per}}

\shorttitle{Properties of the Thick Disk}
\shortauthors{Ruchti et al.}

\begin{document}

\title{Observational Properties of the Metal-Poor Thick Disk of the Milky Way Galaxy and Insights into Its Origins.}

\author{
Gregory~R.~Ruchti\altaffilmark{1,2},
Jon~P.~Fulbright\altaffilmark{1},
Rosemary~F.~G.~Wyse\altaffilmark{1},
Gerard~F.~Gilmore\altaffilmark{3},
Olivier~Bienaym\'{e}\altaffilmark{4},
Joss~Bland-Hawthorn\altaffilmark{5},
Brad~K.~Gibson\altaffilmark{6},
Eva~K.~Grebel\altaffilmark{7},
Amina~Helmi\altaffilmark{8},
Ulisse~Munari\altaffilmark{9},
Julio~F.~Navarro\altaffilmark{10},
Quentin~A.~Parker\altaffilmark{11},
Warren~Reid\altaffilmark{11},
George~M.~Seabroke\altaffilmark{12},
Arnaud~Siebert\altaffilmark{4},
Alessandro~Siviero\altaffilmark{13,14},
Matthias~Steinmetz\altaffilmark{14},
Fred~G.~Watson\altaffilmark{15},
Mary~Williams\altaffilmark{14},
Tomaz~Zwitter\altaffilmark{16,17}
}

\affiliation{
$^1$Johns Hopkins University, 3400 North Charles Street, Baltimore, MD 21218, USA\\
$^2$Current Address: Max Planck Institut f\"ur Astrophyik, Postfach 1317, Karl-Schwarzschild-Str. 1, D-85748 Garching, Germany: gruchti@mpa-garching.mpg.de\\
$^3$Institute of Astronomy, University of Cambridge, Madingley Road, Cambridge CB3 0HA, UK\\
$^4$Observatoire de Strasbourg, 11 Rue de L'Universit\'{e}, 67000 Strasbourg, France\\
$^5$Sydney Institute for Astronomy, School of Physics A28, University of Sydney, NSW 2006, Australia\\
$^6$Jeremiah Horrocks Institute for Astrophysics \& Super-computing, University of Central Lancashire, Preston, PR1 2HE, UK\\
$^7$Astronomisches Rechen-Institut, Zentrum f\"ur Astronomie der Universit\"at Heidelberg, M\"onchhofstr.\ 12--14, 69120 Heidelberg, Germany\\
$^8$Kapteyn Astronomical Institute, University of Groningen, Postbus 800, 9700 AV Groningen, The Netherlands\\
$^9$INAF Osservatorio Astronomico di Padova, Via dell'Osservatorio 8, Asiago I-36012, Italy\\
$^{10}$University of Victoria, P.O. Box 3055, Station CSC, Victoria, BC V8W 3P6, Canada\\
$^{11}$Macquarie University, Sydney, NSW 2109, Australia\\
$^{12}$Mullard Space Science Laboratory, University College London, Holmbury St. Mary, Dorking RH5 6NT, UK\\
$^{13}$Department of Astronomy, Padova University, Vicolo dell'Osservatorio 2, Padova 35122, Italy\\
$^{14}$Leibniz-Institut f\"ur Astrophysik Potsdam (AIP), An der Sternwarte 16, 14482 Potsdam, Germany\\
$^{15}$Australian Astronomical Observatory, Coonabarabran, NSW 2357, Australia\\
$^{16}$Faculty of Mathematics and Physics, University of Ljubljana, Jadranska 19, Ljubljana, Republic of Slovenia\\
$^{17}$Center of Excellence SPACE-SI, Ljubljana, Republic of Slovenia
}

\begin{abstract}

We have undertaken the study of the elemental abundances and kinematic
properties of a metal-poor sample of candidate thick-disk stars
selected from the RAVE spectroscopic survey of bright stars to
differentiate among the present scenarios of the formation of the
thick disk.  In this paper, we report on a sample of 214 red giant
branch, 31 red clump/horizontal branch, and 74 main-sequence/sub-giant
branch metal-poor stars, which serves to augment our previous sample of only giant stars.  We find that the thick disk $\afe$ ratios are
enhanced, and have little variation ($<0.1$~dex), in agreement with our previous study.  The augmented sample further allows, for the first time, investigation of the gradients in the metal-poor thick disk.  For stars with $\feh<-1.2$, the thick disk shows very small gradients, $<0.03\pm0.02~{\rm
dex~kpc^{-1}}$, in $\alpha$-enhancement, while we find a $+0.01\pm0.04~{\rm
dex~kpc^{-1}}$ radial gradient and a $-0.09\pm0.05~{\rm
dex~kpc^{-1}}$ vertical gradient in iron abundance.  In addition, we show that the peak of the distribution of orbital eccentricities for our sample agrees better with models in which the stars that comprise the thick disk were formed primarily in the Galaxy, with direct accretion of stars contributing little. Our results thus disfavor direct accretion of stars from dwarf galaxies into the thick disk as a major contributor to the thick disk population, but cannot discriminate between alternative models for the thick disk, such as those that invoke high-redshift (gas-rich) mergers, heating of a pre-existing thin stellar disk by a minor merger, or efficient radial migration of stars.

\end{abstract}


\section{Introduction}

Since its identification \citep{gilmore83}, the thick disk has been shown
to have distinct kinematics \citep[e.g.][]{chiba00,gilmore02,soubiran03} and a distinct metallicity distribution \citep[e.g.][]{majewski93,chiba00} from the
stellar halo and thin disk of the Milky Way Galaxy.  The formation
history of the thick disk can provide constraints on the origins and formation of the
thin disk and halo, and ultimately the Milky Way Galaxy itself.  The
formation mechanism of the thick disk, however, has been the subject of much discussion for decades.

Soon after the discovery of the thick disk, \citet{jones83} proposed that early star formation in a thin disk, which formed before the Galactic potential reached virial equilibrium, could be subsequently flattened by violent relaxation to form the thick disk.  Later, \citet{burkert92} showed that a thick disk could be formed as the result of a rapid, dissipational collapse accompanied by star formation.

The first model to gain traction, however, was that in which a pre-existing thin stellar disk is heated by a merger with a fairly robust and massive satellite, around $\sim20\%$ the mass of the disk \citep[e.g.,][]{quinn93,velazquez99}.  The thick disk is then formed primarily from the stars in the `thickened' disk, while some stars are directly accreted from the satellite galaxy.  More recently, this model was extended to include cosmologically motivated heating by accretion and merging from many satellites \citep{hayashi06,kazantzidis08,villalobos08}, however, the most massive satellites still dominate the heating.

Another model has the thick disk forming from the direct accretion of stars from satellite galaxies \citep{abadi03}.  Alternatively, the thick disk could have formed during a period of rapid star formation associated with multiple minor, gas-rich mergers \citep{brook04,brook05}.  In this model, the majority of the thick disk stars are formed from the accreted gas in the primordial disk, with a small minority being directly accreted during the mergers.

Models that have more recently gained traction are those that do not necessarily require a merger history to the formation of the thick disk.  Instead, the thick disk could have formed from stars that have radially migrated outward from the inner disk by resonant scattering by transient spiral arms \citep{sellwood02,schonrich09a,schonrich09b}.  Another model, proposed by \citet{bournaud09}, has the thick disk forming from internal gas clumps in the disk at high redshift. 

What is the best way to distinguish these models?  The distributions of kinematics and metallicity provide some discriminants, and these differ between the models most strongly at the metal-poor end.  While there is clearly not a simple one-to-one relationship between age and metallicity, the most metal-poor stars in general formed early, in every scenario.

Elemental abundance patterns reveal even more detailed information about the
star formation history and chemical evolution of a stellar population.
The ratio of $\alpha$-elemental (e.g., Mg, Si, Ca, and Ti) abundances
to iron abundance is sensitive to the relative numbers
of core-collapse Type II supernovae (SNe~II) and Type Ia supernovae
(SNe~Ia) that have occurred in the past.  The $\alpha$-elements are
primarily synthesized in massive stars and ejected in SNe~II on short
timescales ($\sim10^7$~yr) after the formation of the progenitor star,
with the ratio of $\afe$\footnote{We use bracket notation to
indicate relative abundance ratios of two elements A and B: $[{\rm
A}/{\rm B}]=\log[n({\rm A})/n({\rm B})]-\log[n({\rm A})/n({\rm
B})]_{\odot}$.}  from any one SNe~II depending on the mass of the
progenitor massive star \citep{kobayashi06}.  SNe~Ia, which result
from white dwarf remnants in intermediate mass binary systems with
mass transfer, are major sources of iron (and not $\alpha$-elements)
and occur on longer time scales than SNe~II.  The actual
distribution of explosion timescales depends on the details of the
model for the progenitors of Type~Ia SNe, but the onset is always
later than that of Type~II SNe, and there are always explosions
several Gyr after the initial star formation (e.g. Matteucci et
al.~2009).

A self-enriching system will then show
enhanced $\afe$ ratios (compared to the Sun) early in the
star-formation process, when enrichment is dominated by core-collapse
SNe~II, and the actual enhancement is determined by the initial mass
function (IMF) of massive stars (assuming good mixing of the
interstellar medium, ISM, and a well-sampled IMF).  Stars formed after
significant contributions from SNe~Ia to the ISM ($\sim10^8-10^9$~yr
after the first star formation episode) will have considerably more
iron enrichment, and so they will have decreased $\afe$ ratios
\citep[e.g.,][]{matteucci90,gilmore91,matteucci03}.

The models can therefore be tested using a sample of metal-poor ($\feh<-1$) thick disk stars that probe a volume out to several kiloparsec around the Sun.  High resolution spectroscopic observations of each star is necessary to determine the detailed abundances of the $\alpha$-elements in each star.  We identified for study a sample of candidate metal-poor stars in the thick disk selected from the Radial Velocity Experiment survey
\citep[RAVE,][]{steinmetz06}, and observed this sample with high resolution echelle spectrographs at several facilities around the world.  

In \citet[][hereafter
R10]{ruchti10}, we derived several $\afe$ ratios for a subsample of red giant
branch (RGB) and red clump/horizontal branch (RC/HB) stars from our full
candidate metal-poor thick-disk sample.  We found a significant fraction of these stars ($\sim40\%$) to be
most consistent with the thick disk. The abundances of these
metal-poor thick-disk stars had high $\afe$ ratios, consistent
with rapid star formation, which agrees with previous studies of much
smaller samples of metal-poor thick-disk stars \citep{fulbright02,
bensby03, brewer06, reddy06, reddy08,alves-brito10}.  Further, we found that the
$\afe$ ratios were indistinguishable from those of the halo, indicating that the
halo and the thick disk shared a similar massive star IMF and similar efficient mixing of enriched material into the interstellar medium.  

The high $\afe$ ratios of the metal-poor thick-disk giants provided constraints on those models of the formation of the thick disk that are driven by direct accretion of stars from satellite galaxies.  Stars in present-day Milky Way satellites have lower values of $\afe$ at a given $\feh$ (for $\feh\geq-2$) than the stars of the stellar halo, thick disk, or thin disk \citep[see][]{tolstoy09}.  This is understood in terms of the differences in the star formation histories of the stellar populations (as described above), with the dwarf galaxies having very slow enrichment, and strongly non-monotonic star formation rates.  In the models that include direct accretion of stars, accretion takes place well beyond 1~Gyr after the initial star formation episode.  This implies that if each satellite had star formation rates similar to surviving dwarfs, then the accreted stars would have formed after significant iron contribution from SNe~Ia, and thus would have low $\afe$ ratios.  From this we concluded that dwarf galaxies similar to present-day dwarf galaxies
did not play a major role in the formation of the thick disk.  

Further analysis, however, is needed to distinguish those models that do not include a significant fraction of accreted stars from satellite galaxies.  These models all have early, rapid star formation, consistent with high $\afe$ ratios.  A useful diagnostic is the amplitude of the radial and vertical abundance gradients predicted by the models.  For example, the gas-rich merger model predicts more uniform chemical abundances, while low-amplitude vertical gradients are possible in the heating scenario.  Slow, dissipational settling, on the other hand, would produce a significant vertical gradient in metallicity, as well as a possible vertical gradient in the $\afe$ ratios.

In this work, we extend the analysis from R10 to our entire
sample of metal-poor thick-disk candidates.  We further augment that study with a detailed analysis of the abundance gradients in the thick disk. In \S2
and \S3, we briefly describe the candidate selection and high
resolution spectroscopic observations.  In \S4 we describe our stellar parameter
analysis.  We give a full description of our distance estimation 
techniques in \S5, and briefly discuss our population assignments in
\S6 (a full description can be found in R10).  In \S7 we report on the abundance correlations and gradients seen in the data for our full sample, and show that our
conclusions from R10 have not changed by including the dwarfs and
sub-giants in our sample.  We further quantify our abundance results
using IMF-weighted yields in \S9.  We then use orbital eccentricities
of our sample stars in \S10 to distinguish further the models of the
formation of the thick disk.  Finally, we discuss our findings in \S11
and conclude in \S12.

\section{Selection of Candidate Stars for Study}
\label{sec-cand}

As in R10, all candidate stars were first selected from the internal RAVE catalog (Version DR2) to have calibrated metallicities ${\rm [M/H]_{\rm cal}}<-0.7$ \citep[for more details on the
calibration, see][]{zwitter08} and $\teff$ values between 4000~K
and 6500~K.  The low temperature cut was applied to reduce
contamination from metal-rich giant stars and M-dwarfs.  Stars hotter than
6500~K have a higher likelihood of being rapid rotators and may also
have larger non-LTE effects, which could affect our elemental
abundance analysis (as described below).  Those stars that met these parameter constraints and had the highest probability of being
thick disk stars according to their kinematics (based on the RAVE parameters and our distance
estimates, for which the procedure is described below) were selected
for high resolution spectroscopic observations.

\section{High-resolution Echelle Observations}
\label{sec-obs}

Observations were conducted during the period between 2007 May and
2009 February.  All high resolution spectra were obtained using one of the
following echelle spectrographs: MIKE \citep{mike} on the
Magellan-Clay telescope at Las Campanas Observatory in Chile, FEROS
\citep{feros} on the MPG 2.2-m telescope at ESO La Silla Observatory
in Chile, UCLES \citep{ucles} on the Anglo-Australian telescope in
Australia, and the ARC echelle spectrograph \citep{arces} on the
Apache Point 3.5-m telescope in New Mexico, this last facility for stars visible from the
northern hemisphere.

The instumental set-ups gave resolving powers between 35,000-45,000, providing complete spectral coverage from \wave{3500} to \wave{9500}, except for
UCLES which had complete spectral coverage between \wave{4460} and
\wave{7270}.  The echelle spectral data were reduced following the
same procedures as described in R10.  The final reduced spectra all
yielded an S/N ratio greater than 100~${\rm pixel}^{-1}$ at
\wave{5000-6000} and a minimum ${\rm S/N}\sim40$ around \wave{4000},
which is sufficient for detailed elemental abundance analysis.

\subsection{The Final Sample}

Our full sample of metal-poor thick-disk candidate stars consists of the sample from R10
(212 RGB stars and 31 RC/HB stars) and 74 main-sequence or sub-giant
(MS/SG) stars.  An additional 2 RGB stars were discovered during the analysis of the MS/SG stars, bringing our total to 319 candidate metal-poor thick disk stars with high resolution spectroscopic observations.  As indicated in R10, ten of these candidates (5 RGB and 5 MS/SG) were observed twice for internal consistency checks.  Table~\ref{tab-obs} lists the observational data for the sample.

\begin{figure}
\epsscale{1.0}
\plotone{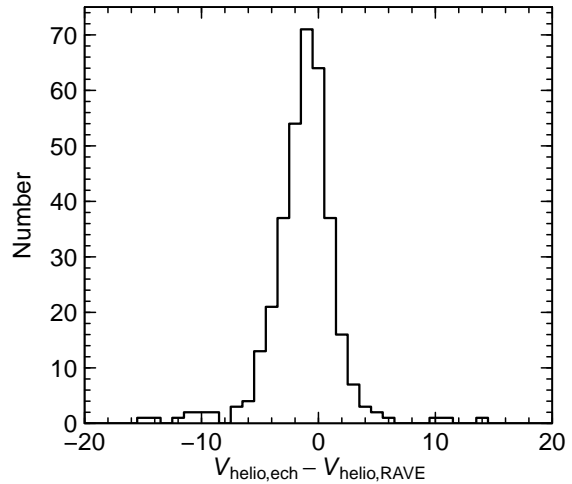}
\caption{Distribution of the difference between the echelle-derived heliocentric radial velocities and those estimated by RAVE.  The average difference is $-1\pm3$~\kmsec.}
\label{fig-vradcomp}
\end{figure}

The full sample has heliocentric radial velocities derived
from the echelle spectra that differ from those in the RAVE database
by an average $-1\pm3$~\kmsec.  Figure~\ref{fig-vradcomp} shows the
distribution of the differences in the radial velocity estimates.  The standard deviation of the RAVE radial velocities is of the order of 2.5~\kmsec, which is slightly lower than that of our comparison.  We cross-checked our final sample with the list of spectroscopic binary stars found in the RAVE survey by \citet{matijevic10}, and found no matches.   Those stars with
differences of 10-15~\kmsec~could possibly be in low-amplitude,
single-lined binary systems in which the primary to secondary ratio is
about 5-10 to one.  This type of binary system would not have a
significant effect on our subsequent analysis and results and we retained these stars.

\begin{figure}
\epsscale{1.0}
\plotone{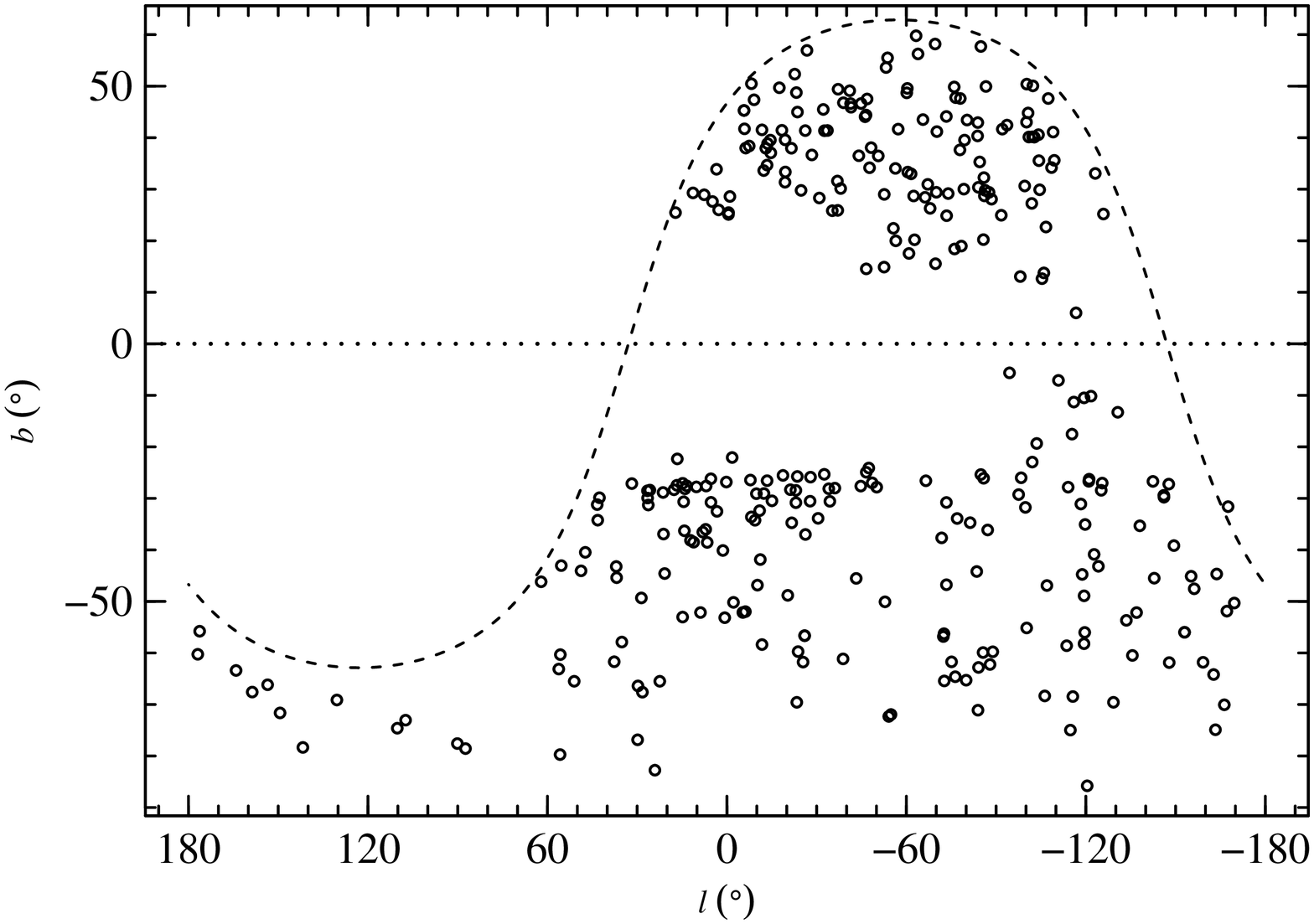}
\caption{Sample coverage in Galactic $(l, b)$ coordinates.  The dotted line represents the Galactic plane, while the dashed curve illustrates the position of the celestial equator in Galactic coordinates.}
\label{fig-coord}
\end{figure}

The coverage in Galactic $(l, b)$ coordinates of our final sample is illustrated in Figure~\ref{fig-coord}.  RAVE mainly targeted fields at Galactic latitudes greater than $20^{\circ}$ (only targeting a few low-latitude fields), which is why there are very few stars with $|b|<20^{\circ}$ in our sample.

\begin{figure}
\epsscale{1.0}
\plotone{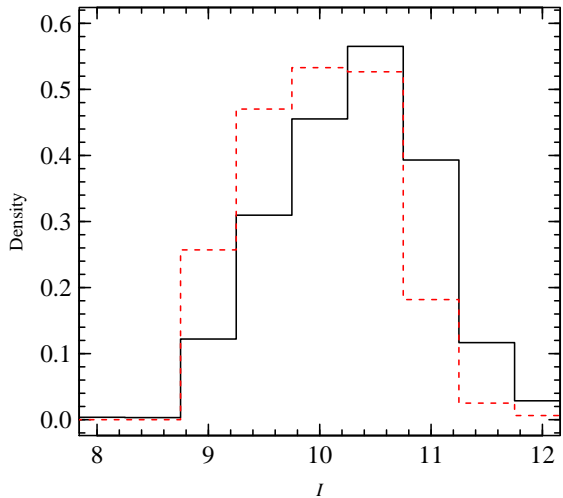}
\caption{Distribution of $I$-magnitude for our candidate sample (red-dashed) and the parent RAVE sample (solid-black).  Note that the RAVE sample only includes those stars that satisfied our constraints in \S\ref{sec-cand}.}
\label{fig-mag}
\end{figure}

 It is clear that our sample of metal-poor thick-disk candidates does
not have the same completeness as the entire RAVE sample from which we selected our candidates.  Figure~\ref{fig-mag} compares
the distribution of the RAVE internal $I$-magnitude of our sample to that of the entire RAVE sample.  Most of our stars lie between $9\leq I\leq11$, about a half magnitude brighter than typical for the parent RAVE sample.  In addition, our selection
function was by no means homogeneous during the observing runs.  It
is important to realize, then, that our sample does not satisfy either
magnitude- or volume-completeness. 

\begin{center}
\begin{deluxetable*}{lrrrccc}
\tablecolumns{7}
\tabletypesize{\scriptsize}
\tablewidth{0pc}
\tablecaption{Observational Data}
\tablehead{
\colhead{Star} & \colhead{RA\tablenotemark{\it a}} & \colhead{DEC\tablenotemark{\it a}} & \colhead{$I$} & \colhead{Obsdate\tablenotemark{\it b,c}} & \colhead{Observatory} & \colhead{S/N\tablenotemark{\it d}} \\
& \colhead{($^{\circ}$)} & \colhead{($^{\circ}$)} & & \colhead{(yyyymmdd)} & & \colhead{(${\rm pix}^{-1}$)}
}
\startdata
C0023306-163143 & 5.878 & -16.529 & 11.2 & 20081015 & LCO & 160 \\
C0315358-094743 & 48.899 & -9.796 & 9.9 & 20081016 & LCO & 170 \\
C0408404-462531 & 62.169 & -46.425 & 11.9 & 20081016 & LCO & 190 \\
C0549576-334007 & 87.490 & -33.669 & 11.0 & 20081015 & LCO & 170 \\
C1141088-453528 & 175.287 & -45.591 & 10.3 & 20070506 & LCO & 100 \\
\enddata
\label{tab-obs}
\tablecomments{Table \ref{tab-obs} is published in its entirety in the electronic edition of the {\it Astrophysical Journal}. A portion is shown here for guidance regarding its form and content.}
\tablenote{equinox 2000}
\tablenote{2009combo or 2007combo refers to the combined spectra from two different nights of the same run.}
\tablenote{For APO runs, `combo' refers to combined spectra from different runs.}
\tablenote{Estimated between \wave{5500-6000}.}
\end{deluxetable*}
\end{center}
 
\section{Abundance Analysis}
\label{parameters}

Initial estimates of the stellar parameters ($\tini$, $\loggi$, $\fehi$) for each star were determined following the
methodology of R10, based on the methods of \citet{fulbright00}.  Full details of this technique can be found in R10.  It is important to note, however, that this method computes the initial quantities using one-dimensional, plane-parallel LTE Kurucz model atmospheres\footnote{http://kurucz.harvard.edu/}.

It is important that we derive accurate stellar parameters since we later use these parameters to determine distances to our stars.  Parameters derived by our initial analysis, however, can often show large differences from expected values computed by different methods.  For example, \citet{ivans01} show, using giants analyzed in the relatively metal-poor ($\feh\sim-1.2$) globular cluster M5, that the $\logg$ derived by methods similar to our initial analysis is often too low, especially for stars near the RGB-tip.  This is in part due to the fact that \nion{Fe}{I} (and somewhat \nion{Fe}{II}) can be strongly affected by non-LTE effects, and so the estimate of surface gravity and iron abundance can be quite unreliable, especially in low-gravity giants and metal-poor stars when ionization equilibrium between \nion{Fe}{I} and \nion{Fe}{II} is assumed in our initial analysis \citep[][Lind, K. \& Bergemann, M., private communication]{thevenin99,asplund99,mashonkina11}.  As summarized in R10, we therefore utilized high-resolution spectroscopic data for several globular cluster stars, as well as reanalyzed several F00 RGB stars with good {\it Hipparcos} parallaxes, to test the accuracy of our initial stellar parameter analysis (derived from our echelle-based analysis) for the giant stars (RGB and RC/HB) in our sample.  Using independent estimates of the stellar parameters for these test cases, we found that the estimates from our initial technique must be systematically corrected to achieve improved accuracy.  Using a sample of 28 F00 MS/SG stars, we have also found that the initial parameters for our MS/SG stars must be similarly corrected.   In the following sections, we give a full description of the corrections to the stellar parameters for our entire sample.

\subsection{Giant Star Parameters}

In R10, we provided a brief summary of the corrections needed for the giant stars in our sample.  In this section, we give the full description of those corrections.  It is important to note that the giant star test samples did not include RC/HB stars, however, since the stellar parameter corrections are independent of the evolutionary state of the giant star, it was assumed in R10 that the RC/HB stars would follow the same corrections as the RGB stars.

\begin{figure}[h]
\epsscale{1.1}
\plotone{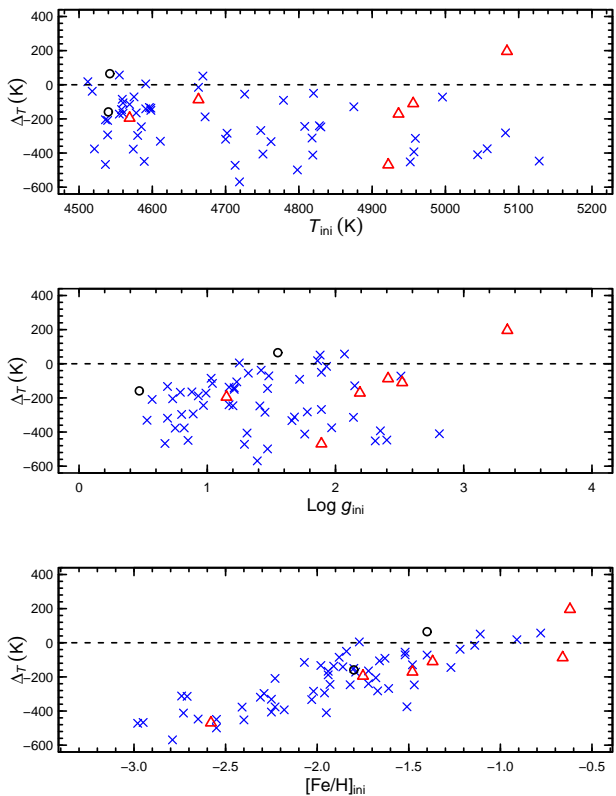}
\caption{$\Delta_T$ vs. $\tini$, $\loggi$, and $\fehi$ for giant stars with $\tini>4500$~K.  Black circles and red triangles represent the globular cluster stars and F00 RGB stars, respectively.  The blue x-symbols are giants from our sample that have small initial reddening ($A_J<0.05$).  Notice that there is no obvious trend with $\tini$ or $\loggi$, however, it is clear that $\tini-\tphot$ depends on $\fehi$.}
\label{fig-gt1}
\end{figure}

\begin{figure}[h]
\epsscale{1.1}
\plotone{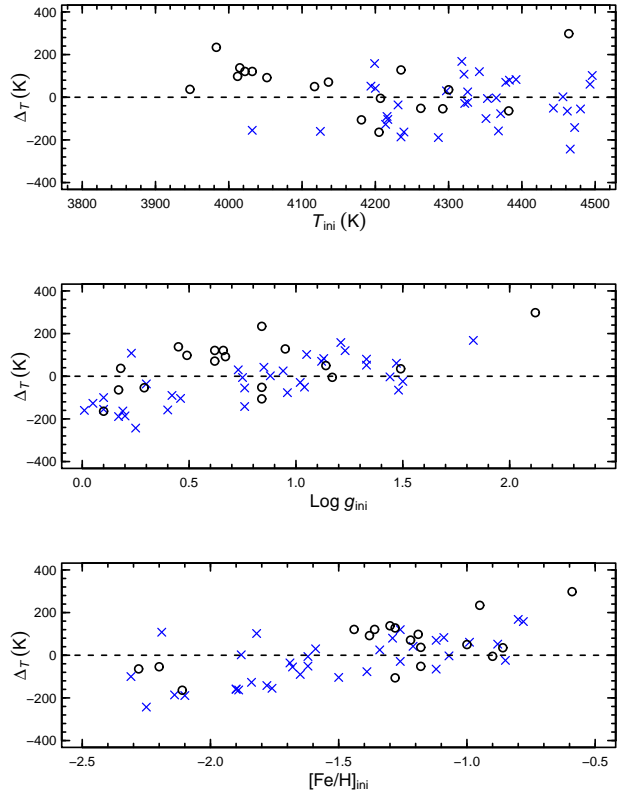}
\caption{Same as Figure~\ref{fig-gt1}, but only for stars with $\tini\leq4500$~K.  Similarly to Figure~\ref{fig-gt1}, the only obvious correlation is with $\fehi$.  It is important to notice, however, that the slope of the correlation is shallower than for the stars with $\tini>4500$~K.  This is the reason we separate the stars into two temperature regimes.}
\label{fig-gt2}
\end{figure}

We computed a photometric temperature ($\tphot$), using the \citet{ghernandez09} 2MASS color-temperature transformations, and a `bolometric' gravity ($g_{\rm bol}=4\pi GM\sigma T^4/L$) for each globular cluster and F00 RGB star.  A photometric temperature was also computed for the stars in our sample that showed low initial reddening ($A_J<0.05$, see \S\ref{sec-red}).  We found in R10 that the offset, $\Delta_T=\tini-\tphot$, shows  different correlations with $\fehi$ depending on the initial $\tini$ of the star.  The behavior can be well approximated by two regimes, hotter and cooler than 4500~K, with correlations as illustrated in Figures \ref{fig-gt1} and \ref{fig-gt2}.

A robust least-squares linear fit to the stars with $\tini>4500$~K (see Figure~\ref{fig-gt1}) resulted in:
\begin{equation}
\Delta_T = \left\{ 
\begin{array}{l l}
0 & \quad {\rm if~\fehi}\geq-1.0\\
240~\fehi + 240 & \quad {\rm if}-1>\fehi \geq-2.5\\
-400 & \quad {\rm if~\fehi}<-2.5\\ 
\end{array} \right.
\label{eq-gt1}
\end{equation}

while for stars with $\tini\leq4500$~K (see Figure~\ref{fig-gt2}):
\begin{equation}
\Delta_T = \left\{ 
\begin{array}{l l}
0 & \quad {\rm if~\fehi}\geq-1.2\\
120~\fehi + 140 & \quad {\rm if~\fehi}<-1.2\\
\end{array} \right.
\label{eq-gt2}
\end{equation}

\begin{figure}[h]
\epsscale{1.1}
\plotone{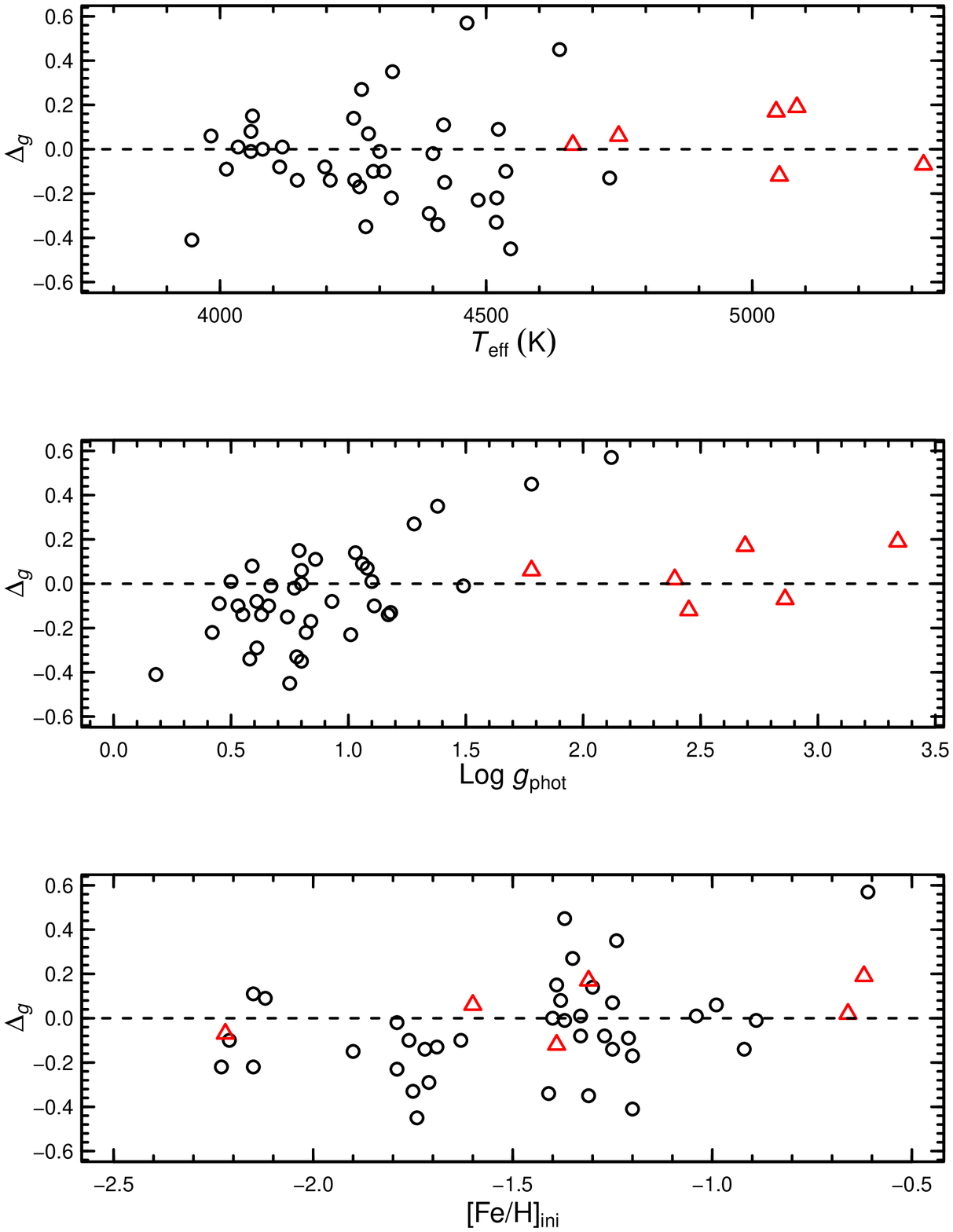}
\caption{Difference, $\loggp-\loggb$, vs. corrected $\teff$, $\loggp$, and $\fehi$.  Black circles and red triangles represent the globular cluster stars and F00 RGB stars, respectively.  Notice that after the temperature correction (equations \ref{eq-gt1} and \ref{eq-gt2}) $\Delta_g$ only depends on $\loggp$.}
\label{fig-gg}
\end{figure}

Following the reasoning of R10, we adopted the above corrections, adding $\Delta_T$ to $\tini$ to obtain final values for the stellar temperature (denoted as $\teff$), to bring our initial temperatures to a scale that will reduce spurious trends as found for our initial spectroscopic analysis.  We then used this final estimate of $\teff$ in the abundance analysis to obtain a new (ionization-balanced) estimate of gravity, denoted $\loggp$.  Comparisons of this $\loggp$ estimate with $\loggb$ showed improvement, but an offset still remained for the lowest gravity stars. This residual offset  correlates with $\loggp$ (see Figure~\ref{fig-gg})  such that a least-squares linear fit to the data resulted in:

\begin{equation}
\Delta_g = \left\{ 
\begin{array}{l l}
0.0 & \quad {\rm if}~\loggp\geq1.0\\ 
0.6~\loggp - 0.6 & \quad {\rm if}~\loggp<1.0\\  
\end{array} \right.
\end{equation}

For those stars with $\loggp<1$, we applied the above correction to the $\loggp$ estimate.  We then adopted both the corrected $\teff$ and this new $\logg$ estimate to get a new estimate of the iron abundance.  We chose the iron abundance from \nion{Fe}{II} as our final estimate of iron abundance to reduce sensitivity to non-LTE effects \citep{thevenin99,asplund99}.  Scatter from our final estimates of temperature and gravity around $\tphot$ and $\loggb$ provided error estimates, $\sigma_{\teff}=140$~K and $\sigma_{\logg}=0.2$~dex.  The iron abundances in the literature are not on a uniform scale, so we estimated the error $\sigma_{\feh}=0.1$~dex, from star-to-star scatter within any one globular cluster.

\subsection{MS/SG Star Parameters}

Only the giant stars in our sample were tested in R10.  We therefore reanalyzed 28 MS/SG stars from F00 to test the accuracy of the initial stellar parameters for the MS/SG stars in our sample.  We computed an independent $\loggb$ and photometric temperature, $\tphot$, following the same methods as for the giants.  We also included estimates of photometric temperature for the MS/SG stars in our sample with low initial reddening (as done for the giant stars).  We found, through a linear fit to the data, that the offset, $\Delta_T=\tini$-$\tphot$, correlates with $\fehi$ for these stars such that:
\begin{equation}
\Delta_T = 190~\fehi + 150
\end{equation}

\begin{figure}[h]
\epsscale{1.1}
\plotone{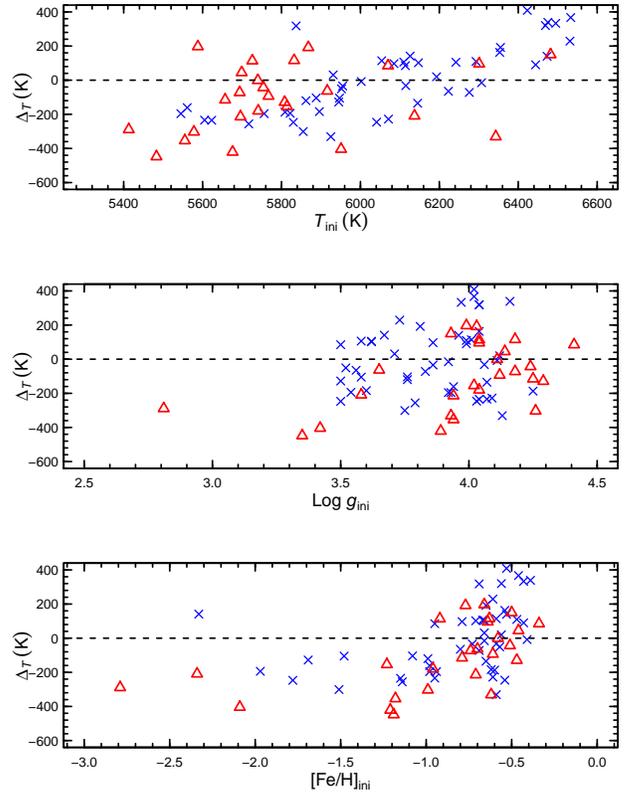}
\caption{Difference, $\tini-\tphot$, vs. $\tini$, $\loggi$, and $\fehi$ for the 28 MS/SG stars from F00.  Red triangles represent the F00 data, while the blue x-symbols represent MS/SG stars in our sample with low initial reddening ($A_J<0.05$).}
\label{fig-dt}
\end{figure}

\begin{figure}[h]
\epsscale{1.1}
\plotone{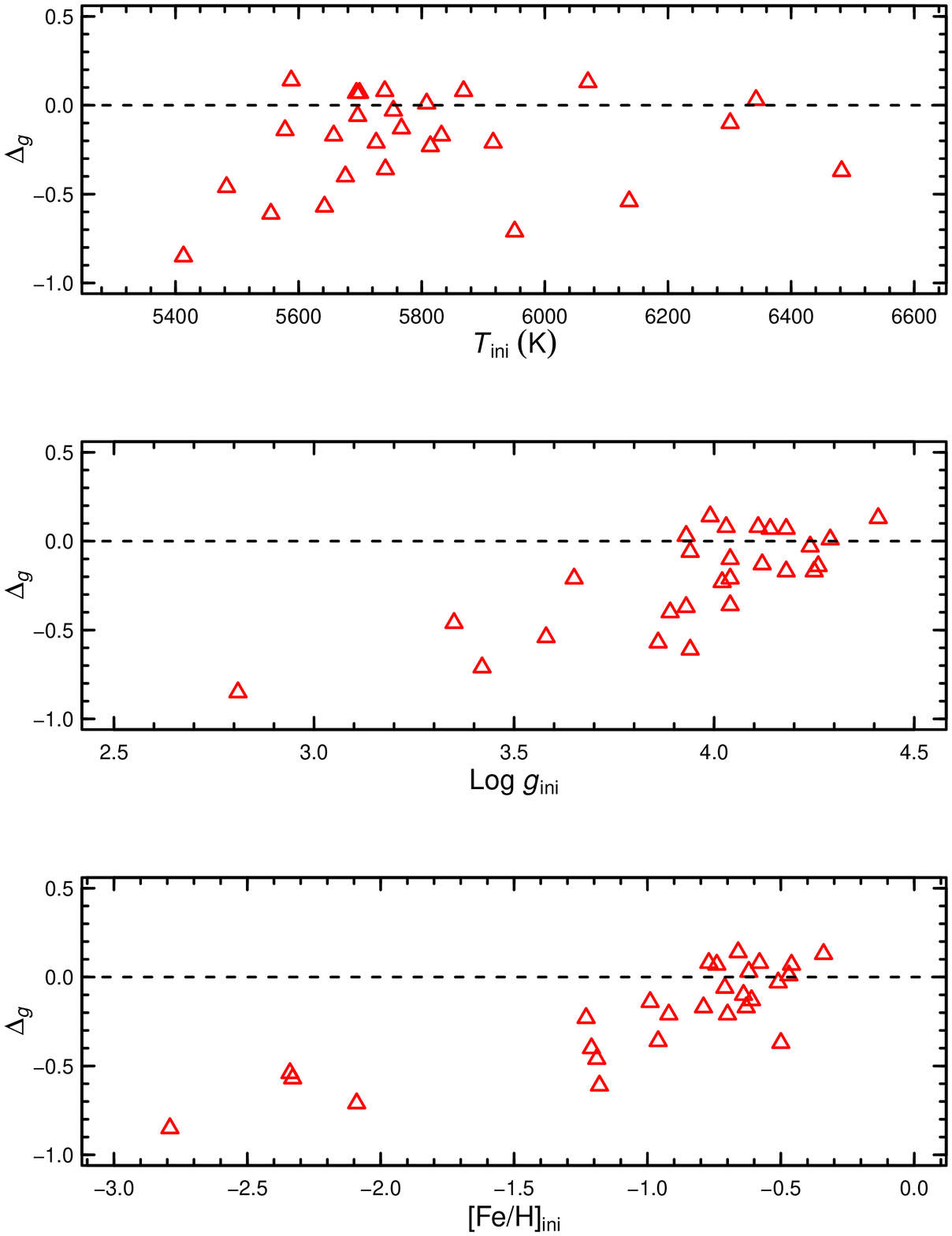}
\caption{$\Delta_g$ vs. corrected $\teff$, $\loggi$, and $\fehi$ for the F00 MS/SG stars.  It is clear from this figure that $\Delta_g$ for the MS/SG stars depends on both $\loggi$ and $\fehi$.}
\label{fig-dg}
\end{figure}

This correlation is illustrated in Figure~\ref{fig-dt}.  In addition, Figure~\ref{fig-dg} shows that the difference, $\Delta_g=\loggi-\loggb$, correlates with both $\fehi$ and $\loggi$ for the MS/SG stars.  The linear fit to this correlation is given by:
\begin{equation}
\Delta_g = 0.3~\fehi + 0.2~\loggi - 0.8.
\end{equation}
We simultaneously corrected our initial $\tini$ and $\loggi$ values according to the above correlations, which resulted in mean scatter around $\tphot$ and $\loggb$ of 150~K and 0.2~dex, respectively.  Again (as for the giants), we chose the abundance of \nion{Fe}{II} as our final estimate of iron abundance.

\subsection{Final Values of Stellar Parameters}

We used the above procedures to ensure that the derived
parameter values for our candidate metal-poor thick disk sample are on the same scales as the globular cluster stars and {\it Hipparcos} stars.  Stars with repeat
observations showed small differences in the values of the stellar
parameters after the corrections above, but they are all smaller
than our estimated errors during the correction procedure.  The effective temperatures showed mean
differences of $16\pm58$~K, the $\logg$ values differed by
$0.05\pm0.12$~dex, and there was a mean difference of
$0.02\pm0.07$~dex in $\feh$.

\begin{figure}[h]
\epsscale{1.1}
\plotone{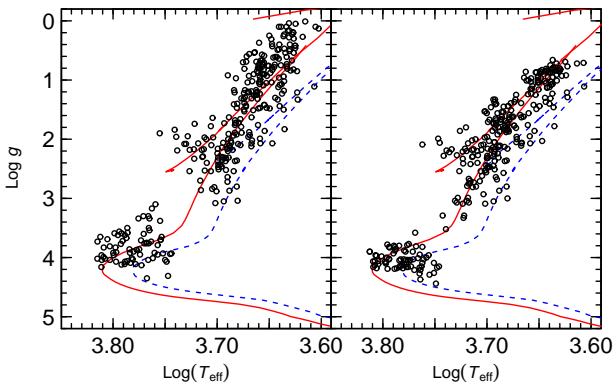}
\caption{Stellar positions in the $\logg$ vs. $\log(\teff)$ plane before (left panel) and after (right panel) our stellar parameter corrections.  The isochrones plotted have an age of 12~Gyr and metallicities $Z=0.001$ (solid red curve) and $Z=0.006$ (dashed blue curve). Some MS/TO stars have different metallicities than the reference isochrones, which makes them appear to be in an impossible part of the plane.  See Figure~\ref{fig-dwarf} for plots of isochrones more appropriate for the MS/SG stars.}
\label{fig-dfit}
\end{figure}

Table~\ref{tab-par} gives our final stellar parameter values for our sample.
Figure~\ref{fig-dfit} illustrates the position of our stars on the
$\logg$ versus $\log(\teff)$ plane before (left panel) and after (right
panel) the above corrections.  Two isochrones, computed by the Padova group \citep{girardi02,marigo08}, with an age of 12~Gyr and
metallicities $Z=0.001$ and $Z=0.006$ are plotted for reference. Notice that the
resulting gravities for the RGB trace the isochrones much better after
the corrections.  Note that some MS/TO stars have different metallicities than the reference isochrones, which makes them appear to be in an impossible part of the plane.

\begin{figure}[h]
\epsscale{1.1}
\plotone{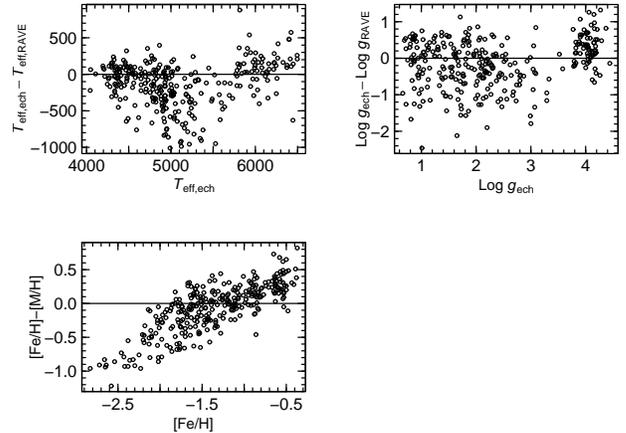}
\caption{Stellar parameter comparisons with RAVE values: $\teff$ (upper left panel), $\logg$ (upper right panel), and metallicity (lower left panel).  All panels plot the echelle value of the parameter minus the analogous RAVE parameter versus the echelle value.  Note that there is a tight correlation between the difference in metallicity values versus the echelle-derived $\feh$, which results from the fact that the RAVE [M/H] is the estimated total metallicity from iron and the alpha elements together.  The more metal-poor stars are more enhanced in the alpha-elements (shown later), which increases this total metallicity.}
\label{fig-raco}
\end{figure}

In Figure~\ref{fig-raco}, we compare our stellar parameter values with those from RAVE.  As can be seen in the figure, on average the RAVE values for $\teff$ and $\logg$ match the echelle-derived values.  There is, however, a large spread, especially for RGB stars at low ($\logg<2$) gravity.  The echelle-derived values also tend to have hotter temperatures and larger gravities for the dwarfs and sub-giants.  Metallicity comparisons show a tight trend between the difference of the two measurements with the echelle-derived value.  It is clear that the RAVE [M/H] is not the same as our echelle-derived $\feh$.

\subsection{Abundances of the Alpha-Elements}

Elemental abundances of several $\alpha$-elements were taken from R10 for the RGB and RC/HB stars, while those for the MS/SG stars were derived using the MOOG analysis program \citep{sneden73} using the final stellar parameters for each star (as done for the giants in R10).  The ratio of the abundance of each $\alpha$-element to iron abundance, 
computed using solar abundances from \citet{grevesse96}, can be found
in Table~\ref{tab-par}.  Comparisons of repeat observations for each ratio, showed differences of $\simlt0.03\pm0.1$~dex.  We
therefore adopted an error of $0.1$~dex in the $\afe$ ratios.
Final abundance values for stars with repeats were taken as the
average of the two estimates for each element.

\begin{center}
\begin{deluxetable*}{lrrrrrrrrr}
\tablecolumns{9}
\tabletypesize{\scriptsize}
\tablewidth{0pc}
\tablecaption{Atmospheric and Abundance-Ratio Data}
\tablehead{
\colhead{Star} & \colhead{$\teff$~(K)} & \colhead{$\logg$} & \colhead{$\feh$\tablenotemark{\it a}} & \colhead{$v_{t}$} & \colhead{Mg\tablenotemark{\it b}} & \colhead{Si\tablenotemark{\it b}} & \colhead{Ca\tablenotemark{\it b}} & \colhead{\nion{Ti}{I}\tablenotemark{\it b}} & \colhead{\nion{Ti}{II}\tablenotemark{\it c}}  \\
}
\startdata
C0023306-163143 & 5528 & 3.29 & -2.30 & 1.4 & 0.32 & 0.46 & 0.34 & 0.11 & 0.15 \\
C0315358-094743 & 4722 & 1.77 & -1.35 & 1.6 & 0.30 & 0.45 & 0.28 & 0.11 & 0.24 \\
C0408404-462531 & 4600 & 0.86 & -2.13 & 2.1 & 0.42 & 0.44 & 0.32 & 0.09 & 0.18 \\
C0549576-334007 & 5379 & 3.04 & -1.76 & 1.3 & 0.24 & 0.22 & 0.31 & 0.15 & 0.21 \\
C1141088-453528 & 4592 & 1.01 & -2.32 & 1.9 & 0.35 & 0.37 & 0.28 & 0.02 & 0.20 \\
\enddata
\label{tab-par}
\tablecomments{Table \ref{tab-par} is published in its entirety in the electronic edition of the {\it Astrophysical Journal}. A portion is shown here for guidance regarding its form and content.}
\tablenote{given as [\nion{Fe}{II}/H]}
\tablenote{given as [X/\nion{Fe}{I}]}
\tablenote{given as [X/\nion{Fe}{II}]}
\end{deluxetable*}
\end{center}

\section{Stellar Distances}
\label{distances}

For this effort to yield meaningful conclusions, we have to determine meaningful population assignments (see \S\ref{sec-pop}).  Meaningful population assignments require accurate kinematic values to distinguish the thick disk from the halo and thin disk.  The greatest contributor to the uncertainty in the kinematics is the distance uncertainty.  We therefore applied techniques that minimized the uncertainty in the derived distances to our stars.

\subsection{RGB Stars}
\label{sec-drgb}

Distances to the RGB stars were adopted from R10, in which we utilized a method that determines probability
weights for a set of isochrones.  An absolute $\mks$ magnitude of a given star was then determined by matching the parameters of a star to  
the most probable isochrone using the technique described below.

\subsubsection{The Isochrones}

Throughout the analysis of R10, as well as this work, we used the set of isochrones computed by the
Padova group.  There are,
however, many other isochrone sets available.  To test the systematics
resulting from using one set of isochrones over another, we also
utilized a second set of isochrones, the Yonsei-Yale
\citep{yi01,demarque04} isochrones, in the analysis of a small subset of our stars covering the full range of stellar parameters for our entire sample.  
The distance estimates derived from fitting the test data to each
isochrone set resulted in an average difference in distance of about
10\% -- well inside our experimental error (as will be discussed
below).  This implies that isochrone differences will not have a
noticeable effect on our estimated distances.  

The Padova isochrones were chosen for the remainder of our analysis since they
can be simply computed for 2MASS magnitudes, and also include the late stages (HB and AGB) of stellar evolution.
For the RGB distance analysis, we created a grid (without interpolation) of Padova
isochrones with metallicities ranging 
between $Z=0.0001$ and $Z=0.01$ (with solar elemental ratios) with a step-size of 0.0002, and 
ages ranging between $4-12$~Gyr with a step-size of 1~Gyr.

\subsubsection{The Fit Technique}

The technique from R10 fit the stellar parameters of each star to several isochrones.  A series of sequential weighted averages over the isochrones were then performed to find the most probable $\mks$ magnitude for each star.  Note that we do not perform any interpolation between isochrone grid points.  The points along the RGB of each isochrone are approximately uniformly distributed, however, the spacing between points on different isochrones (especially of differing metallicity) can be uneven, which may cause over-sampling or under-sampling of different regions of the entire grid.  We attempted to reduce this effect by restricting the range of metallicities used in the fit and introducing a prior probability to each grid point, as described below.

The parameters used to fit each star were $\teff$, $\logg$, and the $Z$-metallicity of the star.  The $Z$-metallicity was determined by combining $\feh$ and the mean $\afe$ using the prescription of \citet{salaris93},
\begin{equation}
Z=Z_o (0.638\cdot10^{-{\rm[\alpha/Fe]}}+0.362), 
\end{equation}
where $Z_o=Z_{\odot}10^{\feh}$ and $Z_{\odot}=0.019$ is the solar metallicity.  Errors in $\feh$ and $\afe$ were propagated to estimate the errors in $Z$.  Next, the probability distribution of each parameter was assumed to be described by a Gaussian,
\begin{equation}
\mathcal{N}(X_{\star},\sigma_{X_{\star}})~\propto~\exp[-(X-X_{\star})^2/2\sigma_{X_{\star}}^2].
\label{eqn-gaus}
\end{equation}
The sigmas, $\sigma_{X_{\star}}$, are the estimated errors in each
parameter, $X_{\star}=\{\teff,~\logg,~Z\}$, derived for each star in \S\ref{parameters}.  The grid of isochrones were then redefined so that for a single age, the grid is limited to only isochrones with metallicities within $\pm10$ times the error of a star's $Z$-metallicity.

In an effort to reduce cases in which two isochrones of differing ages and metallicities might give the same probability, a prior probability function was also introduced.  This function, computed for each point in the grid of isochrones, was derived from the BaSTI luminosity function tracks \citep{pietrinferni04}.  By this method, evolutionary stages with longer lifetimes will be assigned a larger prior probability than short-lived stages, thus biasing the fit to the longer-lived stages \citep[cf.][]{pont04}.

The probability information was then combined so that for each point on an isochrone of a given age, $i$, and metallicity, $j$, a probability weight was computed by,
\begin{equation}
P_{ij}(X_k) = \Psi_{ij}(X_k)\prod_{X_{\star}} \mathcal{N}(X_{\star},\sigma_{X_{\star}})
\label{eqn:pn}
\end{equation} 
where $\mathcal{N}(X_{\star},\sigma_{X_{\star}})$ is given by equation (\ref{eqn-gaus}),  $\Psi_{ij}(X_k)$ is the prior probability, and $X_k$ equals the set of values, $\{\teff,~\logg,~Z\}$, at the given point, $k$, on the isochrone.

The total probability weight for an isochrone of a
given age and metallicity was determined by summing all
$P_{ij}(X_k)$ over all $k$ points in the isochrone.  Further, the most
probable $\mks$ for the given star when matched to that isochrone was computed by performing the
weighted average,
\begin{equation}
M_{ij}=\frac{\sum_k P_{ij}(X_k) M_{ij}(X_k)}{\sum_k P_{ij}(X_k)}, 
\end{equation}
where $M_{ij}$ is the most probable absolute $\mks$ magnitude of the star if it has the age and metallicity of the given isochrone.  

Next, the weighted average of the estimates, $M_{ij}$, of absolute magnitude was performed over all $j$-metallicities to give the most probable estimate $M_i$ of a given age,
\begin{equation}
M_{i}=\frac{\sum_j P_{ij} M_{ij}}{\sum_j P_{ij}}.
\end{equation}
where $P_i=\sum_j~P_{ij}$ is the total probability weight at $i$-age, summed over all metallicities within the range.  Finally, the estimates of the best absolute magnitude, $M_i$, obtained for isochrone ages of 10, 11, and 12~Gyr, were combined by a weighted average to produce the final estimate of the absolute magnitude $\mks$ for each star (as used in R10).

\subsection{RC/HB Stars}

As with the RGB stars, distances to the RC/HB stars were adopted from R10.  Mass loss along the RGB, which affects the position of the HB, is not well modeled in the Padova isochrones.  The absolute $\mks$ magnitude for each RC/HB star was therefore assumed to be equal to the single HB point on the isochrone of equivalent $Z$-metallicity and ${\rm age}=12$~Gyr.

\subsection{MS/SG Stars}

Distances determined by fitting near the turn-off (TO) on isochrones
are much more sensitive to age determination than on the RGB.  In addition, the fit technique for the RGB stars described above can result in unphysical solutions in this region.  The
percent error in the $\teff$ value of a star is typically much smaller
than that of the $\logg$ values, implying the probability distribution close to
the TO region can be double-peaked.  For example, a star with a
gravity that lies to the right of the TO point on an isochrone will
have a probability peak on the MS-branch as well as one on the
SG-branch.  This will cause any weighted averaging of the isochrones
to choose the most probable fit to be a point not on the isochrone
itself, which is physically impossible.  We, therefore, employed a different
method for finding distances to our MS/SG stars to reduce this effect.

Instead of fitting to isochrones, we estimated the absolute $\mks$ magnitude by assuming a mass, using the definition of effective temperature and adopting a bolometric correction for each MS/SG star, which is given by:
\begin{equation}\label{eq-dbol}
\begin{split}
\mks = & -2.5 \left(\log \frac{\rm M}{\rm M_{\odot}} \right. \\
& \left. + 4 \log \frac{\teff}{T_{\odot}} - \log \frac{g}{g_{\odot}} \right) + M_{{\rm bol},\odot} - BC_{\ks}
\end{split}
\end{equation}
where $T_{\odot}=5770$~K, $\logg_{\odot}=4.44$, $M_{{\rm
bol},\odot}=4.72$ is the absolute bolometric magnitude of the Sun, and
$BC_{\ks}$ are the bolometric corrections derived from
\citet{masana06}, which are applicable for the entire metallicity range of our stars and have propagated errors around $\sim0.05$.  This equation, however, still depends on the mass
of a star.  Investigating the mass ranges around the turn-off on the
isochrones shows that MS/SG stars range in mass between
$\sim0.8-1.0$~\msun~ for ages $4-12$~Gyr.  Adopting a mass of
0.8~\msun~implies that the star is old, while adopting a mass of
1.0~\msun~implies that the star is younger.  Comparing a star's
position in the $\logg$ versus $\teff$ plane to a 4~Gyr and 12~Gyr
isochrone, we can determine a reasonable mass estimate.  This is
described in more detail below.

\subsection{Reddening}
\label{sec-red}

An estimate of the reddening to a given star was computed during the procedure to obtain a distance by the following
method.  The \citet{schlegel98} dust maps and extinction calculator
were initially used to calculate reddening to infinity in a specific
line-of-sight.  \citet{bonifacio00} found that \citet{schlegel98}
overestimate the reddening when $E(B-V)_{Sch}> 0.10$ , and 52 of our candidate
stars meet this criterion.  We therefore adopted their correction for
the extinction:
\begin{equation}
\begin{split}
E(B-&V)_C = \\
& \begin{cases}
E(B-V)_{Sch},~~{\rm if}~E(B-V)_{Sch} \leq 0.10 \\
0.10 + 0.65~[E(B-V)_{Sch} - 0.10],~{\rm otherwise.} \\
\end{cases}
\end{split}
\label{eq-red}
\end{equation}

where $E(B-V)_C$ is the reddening estimate after the correction.
Several stars (especially MS/SG stars) are close enough to lie within the Galactic dust distribution, and thus the reddening should be reduced. Adopting a simple exponential model for the dust, with a scale-height of  
$h = 125$~pc \citep{bonifacio00},  the reddening to a star
at distance $D$ and Galactic longitude $b$ is reduced by a factor $1 -
\exp(-|D~\sin~b|/h)$.  The reddening was recomputed iteratively until
the difference between the current estimate of the distance and the previous one was less
than 2 percent.  \citet{schlegel98} assumed a $R_V=3.1$ extinction
curve to compute extinction coefficients in other passbands.  They do
not, however, derive values for the 2MASS passbands.  We determined
our extinction coefficients by assuming the extinction coefficients
from \citet{mccall04},
\begin{equation}
\begin{array}{l l} 
A_J & =0.819~E(B-V)_{C}\\
\aks & =0.350~E(B-V)_{C}. \\
\end{array}
\end{equation}
The mean values of $A_J$ and $\aks$ for our sample are both less than 0.1, with errors less than 0.1~mag. An error of 0.1~mag in reddening will only cause a $\sim4\%$ shift in distance, which is within our error estimates described below.

\subsection{Comparisons \& Error Analysis}

\begin{figure}[h]
\epsscale{1.1}
\plotone{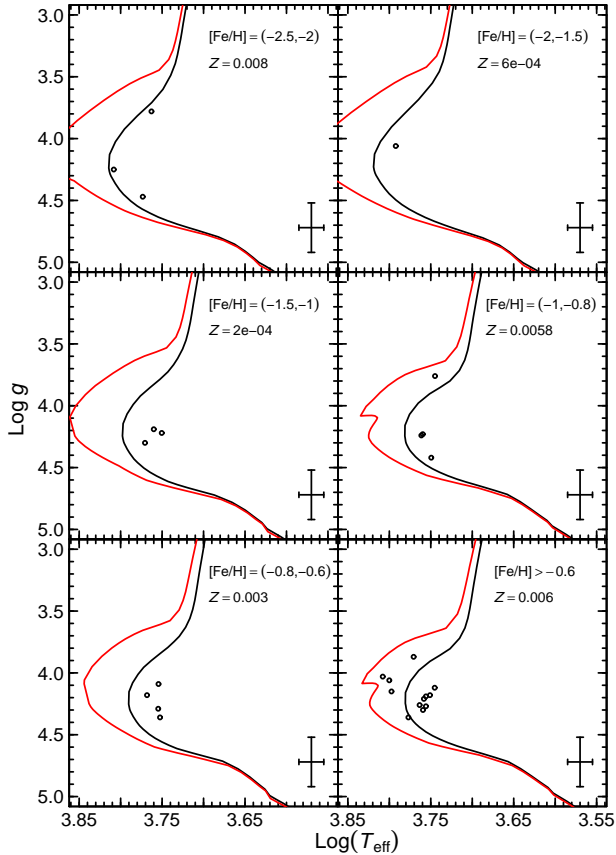}
\caption{Plot of $\logg$ vs. $\log(\teff)$ for the F00 MS/SG stars (after all parameter corrections) for different metallicity bins as shown in each panel.  In addition two isochrone curves are plotted (for the given $Z$-metallicity shown in each panel).  The black curve and red curve represent ages of 12~Gyr and 4~Gyr, respectively.  Note that the majority of the F00 stars appear to be old, except for a few with $\feh>-0.6$.}
\label{fig-hipdwarf}
\end{figure}

We adopted the estimate of 20\% error on the distance to RGB and RC/HB
stars from R10, which was based on comparisons between our estimates and literature estimates for the globular cluster stars and estimates for the F00 RGB stars derived from {\it Hipparcos} parallaxes.  For the MS/SG stars, we applied our technique from above to estimate 
distances for each of the 28 F00 MS/SG stars, and compared to the {\it Hipparcos} distances.
Figure~\ref{fig-hipdwarf} plots these stars in the $\logg$ versus
$\log(\teff)$ plane.  The majority of the stars appear to be old.
There are, however, several stars with $\feh>-0.6$ that could
have younger ages.  If we assume all of the stars are old, and choose
a mass of 0.8~\msun, then the distance estimates from our technique differ from those
derived from the {\it Hipparcos} parallaxes by only $1\%\pm17\%$.
Increasing the mass to 0.9~\msun~and 1.0~\msun~increases the
difference to $7\%\pm18\%$ and $13\%\pm19\%$, respectively.  It is
important to notice that if we increase the mass to one that is
intermediate between older and younger ages, then the difference only
increases by 6\%, which is well within the scatter.

\begin{figure}[h]
\epsscale{1.1}
\plotone{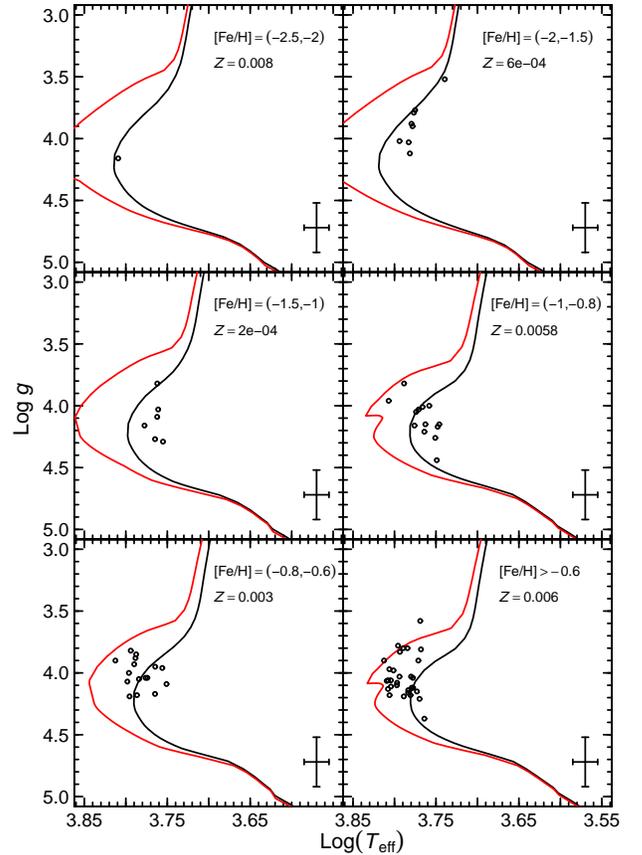}
\caption{Plot of $\logg$ vs. $\log(\teff)$ for the MS/SG stars in our metal-poor sample (after all parameter corrections) for different metallicity bins as shown in each panel.  The isochrones and bins are the same as those in Figure~\ref{fig-hipdwarf}.  Stars with $\feh<-0.8$ are all old, except for two younger stars in the $\feh=(-1,-0.8)$ bin.  We therefore assume a mass of 1.0~\msun~for the two younger stars, while the remaining stars with $\feh<-0.8$ are assumed to have a mass of 0.8~\msun.  The ages of stars with $\feh>-0.8$ are unclear, and cannot be easily separated.  We therefore adopt a mass estimate of 0.9~\msun~for these stars.}
\label{fig-dwarf}
\end{figure}

Figure~\ref{fig-dwarf} shows the same plots as
Figure~\ref{fig-hipdwarf}, except we now plot our metal-poor sample
MS/SG stars.  It is clear that all stars with $\feh<-1$ are
old.  We therefore assume they have a mass of 0.8~\msun.  We assume a
mass of 1.0~\msun~for the two stars with $\feh=(-1,-0.8)$
close to the 4~Gyr isochrone ($\log(\teff)>3.78$), while we adopt
0.8~\msun~for the remaining stars in that bin.  Those stars with $\feh>-0.8$ show no clear separation in age.  We therefore chose an
intermediate mass of 0.9~\msun~for all MS/SG stars with $\feh>-0.8$.  If we apply the above prescriptions to the F00
MS/SG stars, we find a difference between our bolometric distances and
the distances derived from parallaxes to be only $4\%\pm17\%$.  We
therefore adopt a 20\% error on the distances to our MS/SG stars
(similarly to the giants and RC/HB stars), so as to include both
offsets and scatter.

Distance estimates based on RAVE
pipeline values of stellar parameters are also now
available \citep{breddels10,zwitter10,burnett10}. Our distance estimates for the 172
giant stars with $|\logg_{\rm echelle}-\logg_{\rm RAVE}|<0.5$ are shorter
than those of \citet{zwitter10} by $15\%\pm24\%$, while the 73 MS/SG stars have distances shorter by $36\%\pm21\%$.  Note that our technique
was optimized for metal-poor stars and uses parameters derived from echelle spectra,  while
\citet{zwitter10} optimized their method for all stars in the RAVE catalog,
which have a high mean metallicity and typically younger ages and used parameter values from the RAVE pipeline analysis.

\subsection{Final Distances}

\begin{figure}[h]
\epsscale{0.8}
\plotone{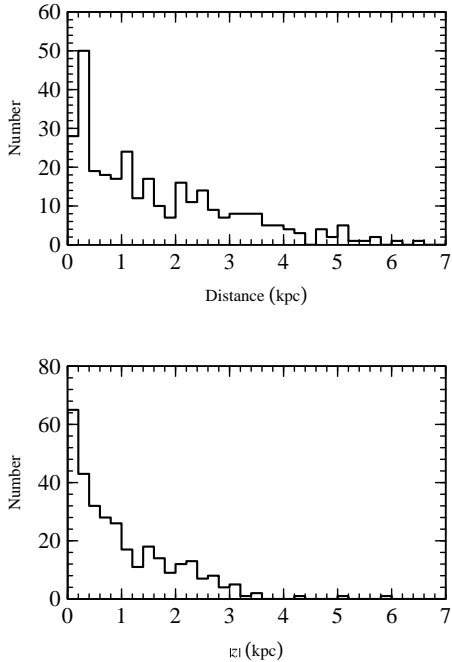}
\caption{Distribution of total distance (upper panel) and vertical $|z|$-height (lower panel) for our candidate sample.  Note that the star at a distance of $\sim16$~kpc is not shown.}
\label{fig-hdist}
\end{figure}

Our final estimate of the distance to each star, from the Sun, can be found in Table~\ref{tab-kin}, with the error being 20\% of the distance estimate. The average distance to the RGB and RC/HB stars is $\sim2$~kpc.  All had distances less than $\sim7$~kpc, except one at $\sim16$~kpc (see Figure~\ref{fig-hdist}).  The MS/SG stars in our sample have an average distance of $\sim220$~pc, extending out to $\sim400$~pc.  The majority of our stars (primarily giants) extend to an average vertical height of $|z|\simgt1$~kpc, as shown in Figure~\ref{fig-hdist}.  It is clear that our sample probes distances much further than the solar neighborhood ($\sim100$~pc) -- not done by any previous analyses of elemental abundances of metal-poor thick disk stars in the literature.

\subsection{Velocities and Orbits}
\label{velorb}

We computed three-dimensional space motions of our stars in
cylindrical coordinates (given in Table~\ref{tab-kin}) by combining
the distances and radial velocities derived from the analysis of our spectra
with the proper motions given in the RAVE database.  We further derived
the full orbit of each star by assuming a Galactic potential and
combining that with the velocity and position information for each
star.  Stellar orbits were computed over 15~Gyr using an orbital integrator based
on a three-component Galactic potential.  The disk is modeled as a
Miyamoto-Nagai potential \citep{miyamoto75}, while we used the
\citet{hernquist90} potential for the bulge.  Finally, we assumed a
logarithmic spherical potential for the halo.  We took $M_{\rm
disk}=8.0~{\rm x}~10^{10}$~\msun, $M_{\rm bulge}=2.5~{\rm
x}~10^{10}$~\msun, with characteristic scales $a=6.5$, $b=0.26$,
$c=0.7$, and $d=12.0$, all in kiloparsecs, and $v^2_{\rm halo}=27000\, ({\rm km~s^ {-1}})^2$.  These values ensure that
the circular velocity equals 220~\kmsec~at 8~kpc from the
Galactic center.  

The orbital parameters of each star are
listed in Table~\ref{tab-orb}. Two sets of data are listed.  The maximum vertical distance, $|z_{max}|$, and the closest and furthest distances,$\rper$ and $\rapo$, reached by a star for all orbits integrated are listed in columns  3 through 5, while the same parameters for the final orbit of the star are listed in columns 6 through 8.  The total number of orbits integrated is given as $N_{\rm orbit}$.  The eccentricity of any given orbit of a star is defined as 
\begin{equation}
\varepsilon = \frac{\rapo - \rper}{\rapo + \rper}.
\end{equation}
The parameters for the last orbit were chosen to compute $\varepsilon$  in order to make direct comparisons of stellar eccentricities with model simulations (see \S\ref{ecc}).  The amplitude of variation for $\rper$ and $\rapo$ depends on the $z$-excursions during each orbit of a star, and thus, can be quite large.  The amplitude of variation in the eccentricity of an orbit, however, is typically less than 0.05.  We therefore assume the eccentricity of the last orbit is representative of the true orbital eccentricity of a star.

\begin{center}
\begin{deluxetable*}{lrrrrrrrrrrrr}
\tablecolumns{13}
\tabletypesize{\scriptsize}
\tablewidth{0pc}
\setlength{\tabcolsep}{0.04in}
\tablecaption{Kinematic Data}
\tablehead{
\colhead{Star} & \colhead{$D$} & \colhead{$V_{\rm helio}$} & \colhead{$V_{\Pi}$} & \colhead{$\epsilon_{\Pi}$} & \colhead{$V_{\Theta}$} & \colhead{$\epsilon_{\Theta}$} & \colhead{$V_{Z}$} & \colhead{$\epsilon_Z$} & \colhead{$P_{\rm thin}$} & \colhead{$P_{\rm thick}$} & \colhead{$P_{\rm halo}$} & \colhead{POP} \\
& \colhead{(pc)} & \colhead{(\kmsec)} & \colhead{(\kmsec)} & \colhead{(\kmsec)} & \colhead{(\kmsec)} & \colhead{(\kmsec)} & \colhead{(\kmsec)} & \colhead{(\kmsec)} & & & }
\startdata
C0023306-163143 & 921 & -7.0 & 53.2 & 19.3 & -195.3 & 85.9 & -77.9 & 19.0 & 0.00 & 0.00 & 1.00 & 3 \\
C0315358-094743 & 2434 & 131.6 & 86.3 & 14.6 & 50.4 & 36.4 & -54.0 & 15.2 & 0.00 & 0.13 & 0.87 & 3 \\
C0408404-462531 & 15699 & 52.3 & 74.2 & 344.8 & -88.9 & 338.4 & 241.7 & 269.9 & 0.00 & 0.01 & 0.99 & 3 \\
C0549576-334007 & 1157 & 79.7 & -29.1 & 16.4 & 134.6 & 11.7 & -24.7 & 14.2 & 0.02 & 0.91 & 0.07 & 2 \\
C1141088-453528 & 6152 & 83.9 & -105.0 & 54.2 & 58.1 & 51.9 & -88.1 & 69.2 & 0.00 & 0.11 & 0.89 & 3 \\
\enddata
\label{tab-kin}
\tablecomments{Table \ref{tab-kin} is published in its entirety in the electronic edition of the {\it Astrophysical Journal}. A portion is shown here for guidance regarding its form and content.}
\end{deluxetable*}
\end{center}

\begin{center}
\begin{deluxetable*}{lrrrrrrrr}
\tablecolumns{7}
\tabletypesize{\scriptsize}
\tablewidth{0pc}
\tablecaption{Orbital Parameters}
\tablehead{
 & & \multicolumn{3}{c}{All Orbits} & \multicolumn{3}{c}{Final Orbit} & \\
\colhead{Star} & \colhead{${\rm N_{orbit}}$} & \colhead{$\rper$} & \colhead{$\rapo$} &  \colhead{$|z_{\rm max}|$} & \colhead{$\rper$} & \colhead{$\rapo$} &  \colhead{$|z_{\rm max}|$} & \colhead{$\varepsilon$} \\
& & \colhead{(kpc)} & \colhead{(kpc)} & \colhead{(kpc)} & \colhead{(kpc)} & \colhead{(kpc)} & \colhead{(kpc)} & 
}
\startdata
C0023306-163143 & 100 & 6.5 & 9.2 & 1.6 & 6.6 & 9.1 & 1.5 & 0.2 \\
C0315358-094743 & 116 & 1.4 & 10.6 & 2.3 & 1.6 & 10.5 & 2.2 & 0.7 \\
C0408404-462531 & 61 & 15.9 & 23.8 & 23.8 & 16.9 & 23.5 & 23.4 & 0.2 \\
C0549576-334007 & 119 & 3.9 & 8.9 & 0.6 & 3.9 & 8.9 & 0.6 & 0.4 \\
C1141088-453528 & 127 & 0.6 & 10.6 & 3.4 & 0.7 & 9.8 & 2.4 & 0.9 \\
\enddata
\label{tab-orb}
\tablecomments{Table \ref{tab-orb} is published in its entirety in the electronic edition of the {\it Astrophysical Journal}. A portion is shown here for guidance regarding its form and content.}
\end{deluxetable*} 
\end{center}

\section {Population Assignments}
\label{sec-pop}

Each star was assigned to a Galactic population following the same Monte-Carlo method as in R10.  We drew 10,000 random samples of each component of a star's space motion from a Gaussian error distribution centered on our estimate of the component velocity.  The probabilities that each random set of velocities was drawn from the thin disk, thick disk, or halo were then computed using the local characteristic Gaussian definitions for each Galactic population (given in Table~1 of R10 and reproduced here in Table~\ref{tab-gaus}).  A random set was assigned to a specific Galactic population if the probability was four times that of the other two probabilities.  A star was then finally assigned to the Galactic population with the most random set assignments.  

It is important to note that some stars had probabilities that did not easily distinguish between the Galactic populations (the ratio of the probabilities of two Galactic components was less than four).  These stars were then assigned to additional intermediate thin/thick or thick/halo populations.  This method, however, is susceptible to possible misassignments.  During the analysis, we therefore retained an additional probability statistic. which equalled the sum of the probabilities (normalized) obtained from all Monte-Carlo realizations for each Galactic population (hereafter, the PDF values).

Similarly, random sets of a star's distance were drawn from a Gaussian error distribution centered on our estimate of the distance and a sigma equal to 20 percent of the distance.  The probability that a random `star' lies in each Galactic population was computed by comparing to the characteristic double-exponential distributions for the thin and thick disks and the two-axial power-law ellipsoid for the halo, taken from \citet{juric08}.  As with the velocities, a star was assigned to the Galactic population with the highest number of occurrences from the re-sampled distance.  The positional assignment was then used as a boundary condition such that if a star was assigned to the thick disk from its velocities and it was assigned to the halo from its position, the star would then be assigned to the halo.  A star would remain assigned to the thick disk, however, if it was assigned to the thin disk from its position.

\begin{deluxetable*}{lccccr}
\tablecolumns{6}
\tabletypesize{\scriptsize}
\tablewidth{0pc}
\tablecaption{Local Characteristic Velocity Distributions}
\tablehead{
\colhead{Population} & \colhead{$\sigma_{\Pi}$} & \colhead{$\sigma_{\Theta}$} & \colhead{$\sigma_{Z}$} & \colhead{$\langle V_{
\Theta} \rangle$} &  \colhead{Ref.} \\
& \colhead{(\kmsec)} & \colhead{(\kmsec)} & \colhead{(\kmsec)} & \colhead{(\kmsec)} & 
}
\startdata
Thin Disk & 39 & 20 & 20 & -15 & \citet{soubiran03} \\
Thick Disk & 63 & 39 & 39 & -51 & \citet{soubiran03}\\
Halo  & 141 & 106 & 94 & -220 & \citet{chiba00}
\enddata
\label{tab-gaus}
\end{deluxetable*}

\begin{figure}
\epsscale{1.1}
\plotone{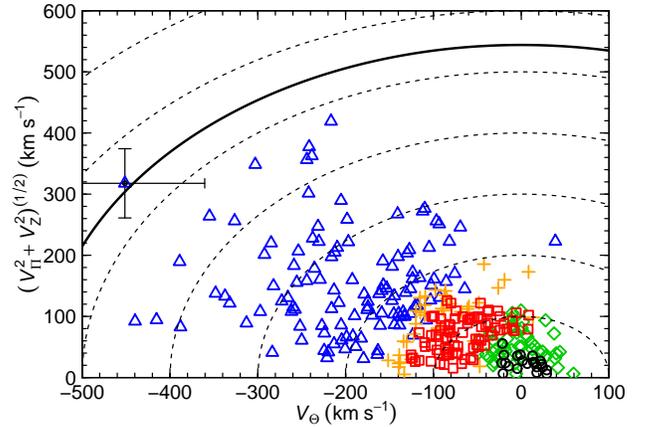}
\caption{Toomre diagram for our full sample with
$\sigma_{V_{\theta}}<100$~\kmsec.  The black circles, green diamonds,
red squares, orange plus signs, and blue triangles correspond to thin
disk, thin/thick, thick disk, thick/halo, and halo stars,
respectively.  The dashed curves indicate constant space motion, while an estimate of the local 
escape velocity, based on radial velocities within the RAVE database \citep{smith07}, is represented by the thick solid curve.  Typical velocity errors are $1\sigma$ error  $<20$~\kmsec.}
\label{fig-toom}
\end{figure}

\begin{figure}[h]
\epsscale{1.1}
\plotone{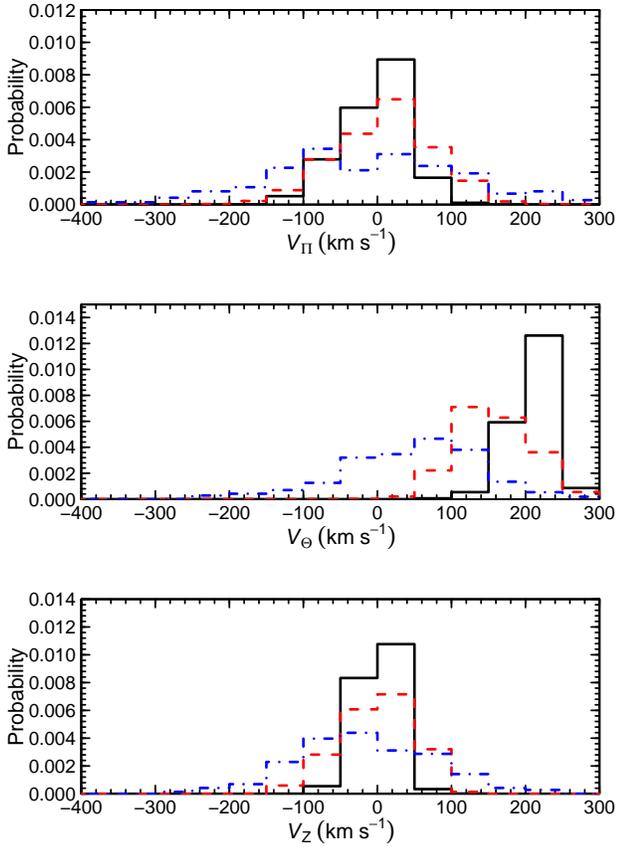}
\caption{Distribution of $V_{\Pi}$ (top panel), $V_{\Theta}$ (middle panel), and $V_Z$ (bottom panel) velocities for our stars given the PDF value that they belong to the thin disk (solid black), thick disk (dashed red), and halo (dot-dashed blue).  Note that each distribution was created by summing the PDF values within a given velocity bin.  The distributions were then normalized such that the total area equals unity.}
\label{fig-dvel}
\end{figure}

The stars were assigned as follows: 88 thick disk, 21 thin disk, 51
intermediate thin/thick disk, 36 intermediate thick/halo, and 123
halo.  The last four columns of Table~\ref{tab-kin} gives the PDF values and Galactic
population assignment for each star.  A Toomre diagram is plotted in Figure~\ref{fig-toom} to illustrate the
relationship between our final population assignments and stellar
velocities.  Note that comparisons with the Toomre diagram from R10 show
that the MS/SG stars mainly comprise the thin disk and thin/thick
population.  We further sum the stars' PDF values for each Galactic population within a given velocity bin and plot the distribution of each velocity component in Figure~\ref{fig-dvel}. It is clear that the distributions reflect the underlying assumed populations.

\section{The Metal-Poor Thick Disk}

\subsection{Iron Abundance Distribution and Gradients}

\begin{figure}[h]
\epsscale{1.1}
\plotone{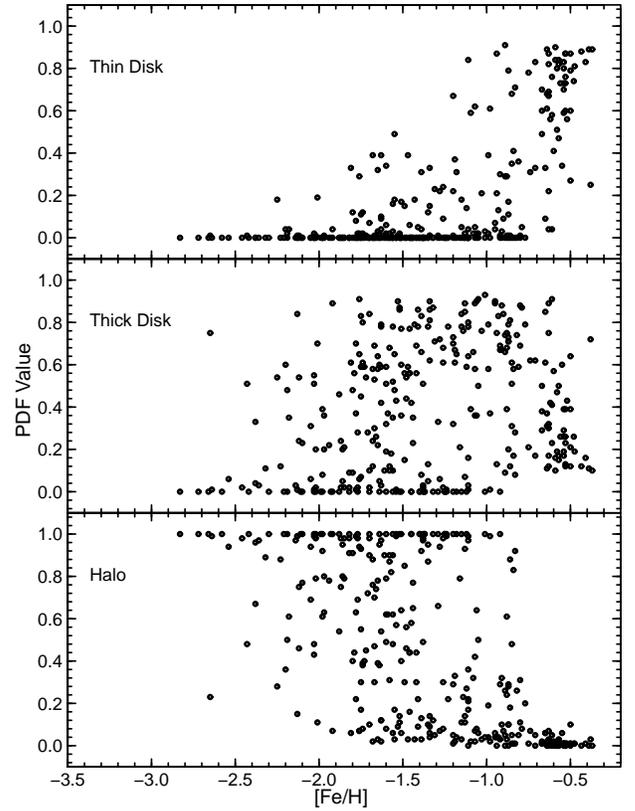}
\caption{PDF values for each Galactic population versus $\feh$.}
\label{fig-zpdfs}
\end{figure}

\begin{figure}[h]
\epsscale{1.1}
\plotone{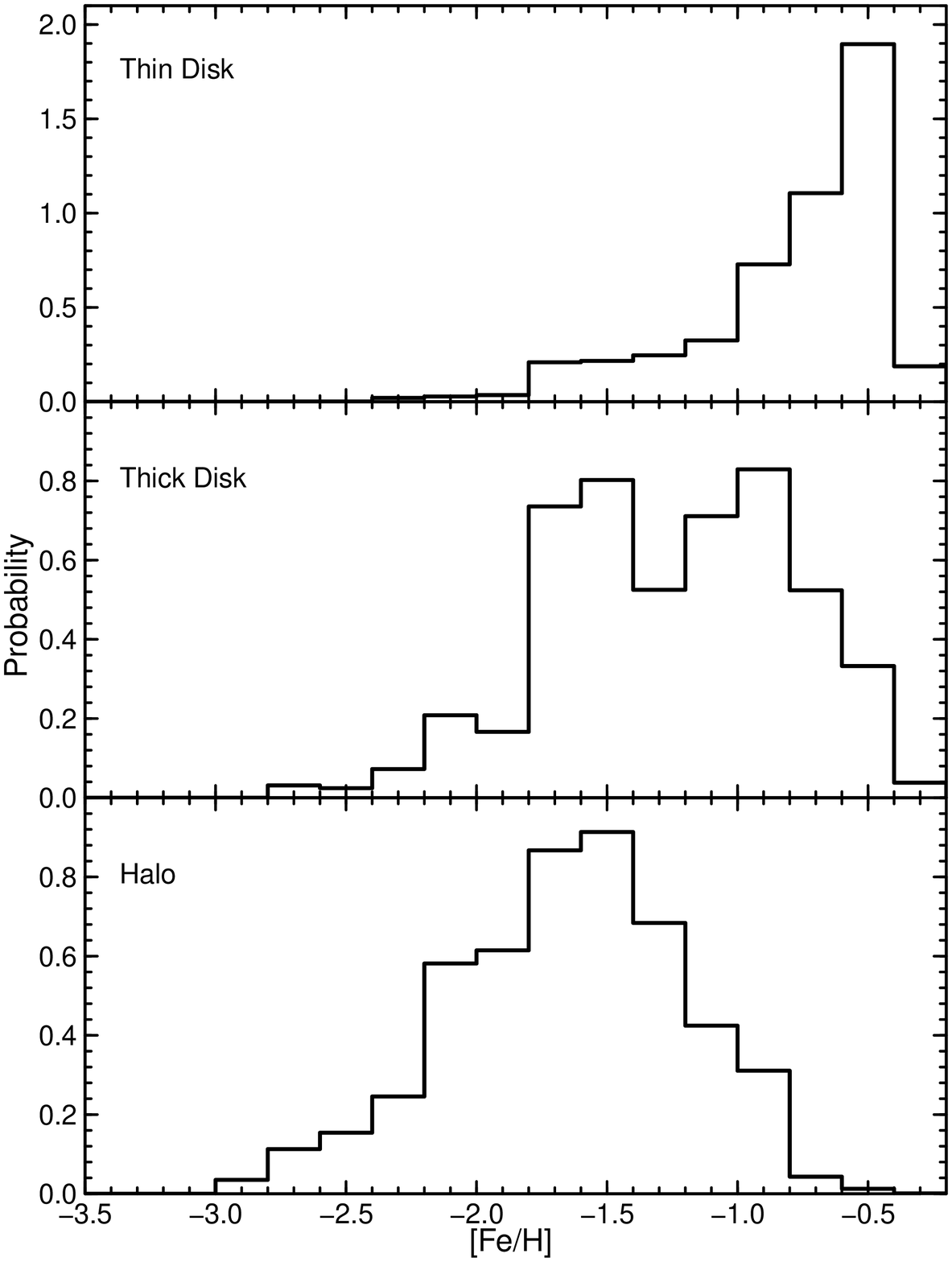}
\caption{Distribution of $\feh$ for thin disk, thick disk, and halo.  Note that each distribution was created by summing the PDF values within a given metallicity bin.  The distributions were then normalized such that the total area equals unity.  Recall that our sample was selected to be metal-poor, so that  the full metallicity distributions of each component are not uniformly represented.}
\label{fig-zdists}
\end{figure}

Recall (from \S\ref{parameters}) that we estimate the iron abundance
of each star from \nion{Fe}{II}, since \nion{Fe}{II} is both the
dominant species and is much less sensitive to non-LTE
effects than is \nion{Fe}{I}
\citep{thevenin99,asplund99,mashonkina11}.  Figure~\ref{fig-zpdfs} shows the PDF values for each Galactic population versus $\feh$, while Figure~\ref{fig-zdists} shows the distribution of iron abundance for each of the three Galactic populations. 
As was found in R10, the majority of the thick disk has $\feh>-1.8$, with a small tail to much lower metallicities, the lowest metallicity being $\feh\sim-2.7$.  The MS/SG stars assigned to the thick disk predominantly have metallicities close to -1~dex, with a short tail down to $\feh\sim-1.8$.  The distribution of $\feh$, therefore did not show a significant change from that of R10.  Due to sample selection
effects, we do not completely sample the high-metallicity ($\feh>-1$) parts of the distributions.  Further, the offset between our final $\feh$ values and the RAVE metallicities (see Figure~\ref{fig-raco}) contributed to the shortage of thick disk stars with metallicities at the peak from local studies ($\feh\sim-0.6$).  Stars that may have seemed to have metallicities near -0.6~dex, would end up with a lower value for the final metallicities.

The most notable attribute of the thick disk metallicity distribution is that it appears to be double-peaked.  The second, low-metallicity peak is comprised of many stars that have intermediate probabilities for both the thick disk and halo (${\rm PDFs}<0.6$, assigned as thick/halo), while the peak above $\feh\sim-1$ is comprised of stars with probabilities of being thick disk equal to $0.6-1.0$ (see Figure~\ref{fig-zpdfs}).  It is therefore important to note that the low-metallicity peak is not just systematic noise in the population assignments, but is actually different.  Could this be a metal-poor `sub-component' of the thick disk, or could it be the high angular momentum tail of the halo?  We will come back to this later (see \S\ref{sec-hianghalo}). 

\begin{figure}[h]
\epsscale{1.1}
\plotone{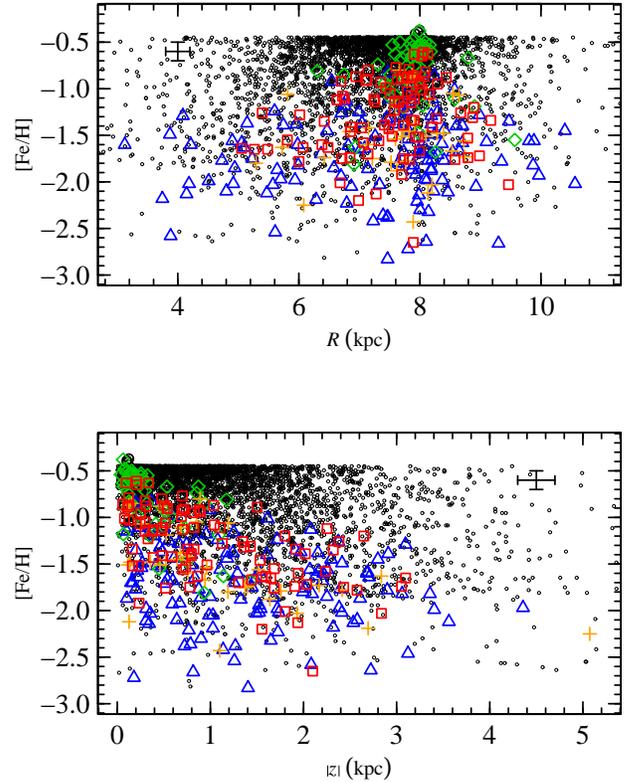}
\caption{$\feh$ vs. $R$ and $|z|$ for our sample of stars (symbols and colors
are the same as Figure~\ref{fig-toom}).  The small data points are
those stars from our original RAVE catalog, from which our sample was
taken.  The RAVE catalog stars populate large $|z|$, even at high
$\feh$, which suggests that the lack of stars in our metal-poor sample with
$\feh>-1$ at large $|z|$ is due to the selection
function of the sample, combined with the subsequent temperature cuts
prior to the abundance analysis.}
\label{fig-zgcomp}
\end{figure}

Before we can determine the presence of iron abundance gradients in
the thick disk, we must first determine possible selection effects in
our sample and introduced through the analysis.  For example, we introduced a temperature cut, such that only stars with temperatures in the range $4000-6500$~K were selected for observation, to reduce contamination of stars that would most likely fail our abundance analysis (see \S\ref{sec-cand}).
Figure~\ref{fig-zgcomp} plots $\feh$ versus $R$ and $|z|$ for our
sample on top of the RAVE catalog, with ${\rm [M/H]}<-0.5$, from which our sample stars were selected.  We performed a linear fit to the $\feh$ vs. $\feh-\mh$ relation in Figure~\ref{fig-raco} in order to put our sample stars and the RAVE catalog stars on the same scale.  Distances to the RAVE catalog stars were
computed using an analogous method to that described in
\S\ref{distances}.  Our sample does not show the same
distance coverage as the RAVE catalog stars for $\feh\geq-1.2$.
This implies that the lack of thick-disk stars with
$\feh\geq-1.2$ at large radial and vertical distances is
possibly an artifact of our cuts in temperature.

\begin{figure}[h]
\epsscale{1.1}
\plotone{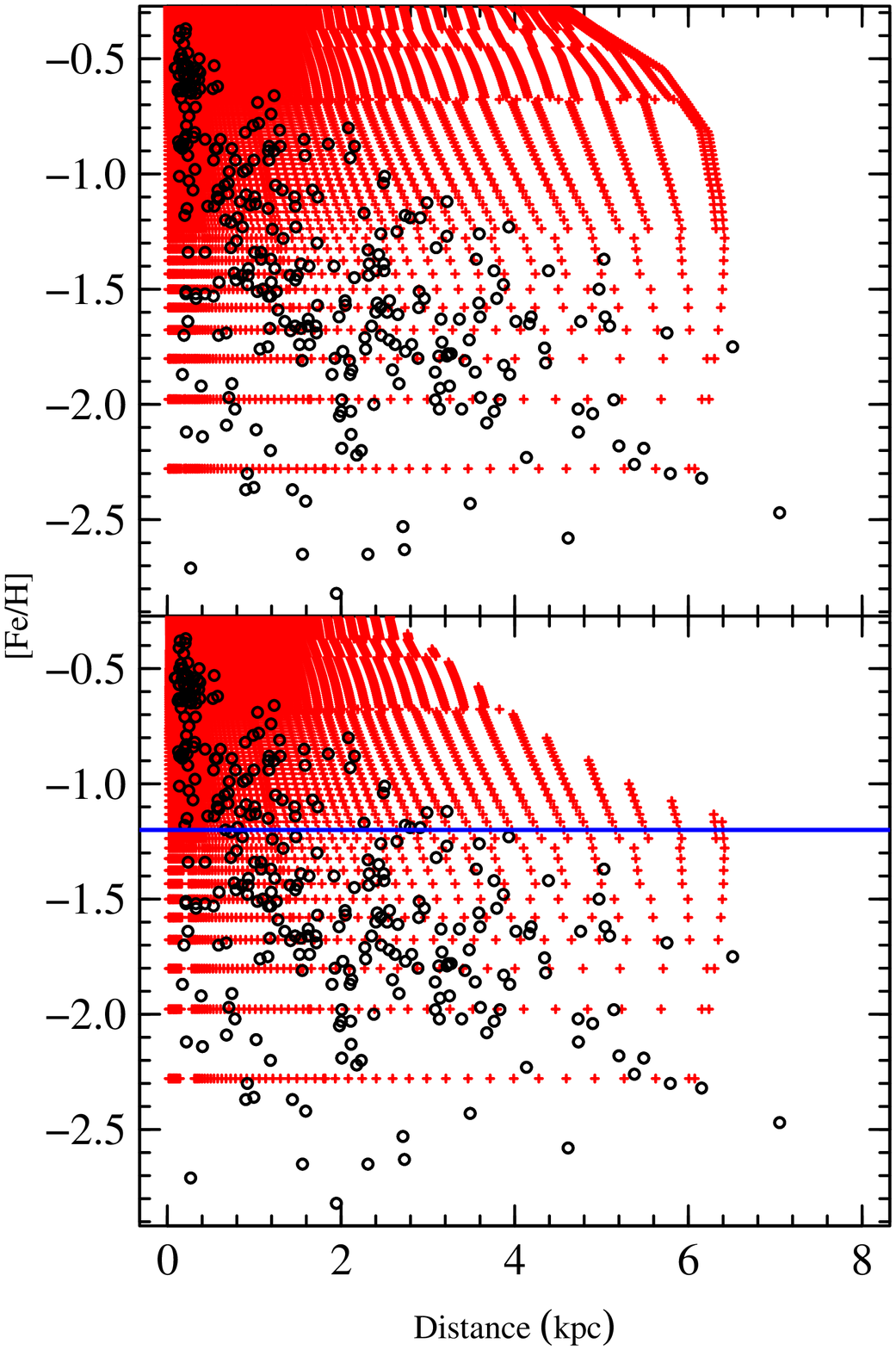}
\caption{$\feh$ vs. Distance for our sample of stars.  The red `plus' symbols represent a 12~Gyr isochrone of varying metallicity, for which distances are computed assuming an apparent magnitude of $I=10$.  The black points represent our sample of candidate metal-poor thick disk stars. Note that there is one star not shown at a distance of 15~kpc with $\feh\sim-2.2$.  The top panel shows the isochrone points without any temperature cut, while the same temperature cut as that of our candidate sample ($4000-6500$~K) is applied to the isochrone data in the bottom panel.  The brightest metal-rich giants are cooler than their metal-poor counterparts, and so the low-temperature limit cuts them out and makes the difference in isochrones seen in the two panels.  For $\feh<-1.2$ (blue line) the sample is unaffected by our temperature cut.}
\label{fig-gbias}
\end{figure}

We investigated the possible origin of these selection
effects by comparing to an old (12~Gyr) isochrone.  In
Figure~\ref{fig-gbias}, we plot the derived data for our sample together with 12~Gyr
isochrones of differing metallicities.  The metallicities of the isochrones were simply converted to $\feh$ by $\feh=\log({\it Z/Z}_{\odot})$ as a first-order approximation.  We applied an apparent
magnitude limit of $I=10$, the peak $I$-magnitude of our sample (see Figure~\ref{fig-mag}), to the isochrone data to determine the
maximum distance that can be observed for each metallicity.  It is
clear that the stars of our sample do not reach the maximum limit (as shown in
the top panel of Figure~\ref{fig-gbias}).  However, when we apply the
same temperature cut  that we applied during our candidate selection
($4000-6500$~K) to the isochrone data, then we see a similar trend emerge in the isochrone behavior as in the sample (lower panel, Figure~\ref{fig-gbias}).  This is a clear illustration
of the selection effects in our final sample.  For $\feh<-1.2$,
however, the isochrone data are unaffected by the temperature cut, suggesting that below this value the data for our sample may be used without further corrections.

\begin{figure}[h]
\epsscale{1.1}
\plotone{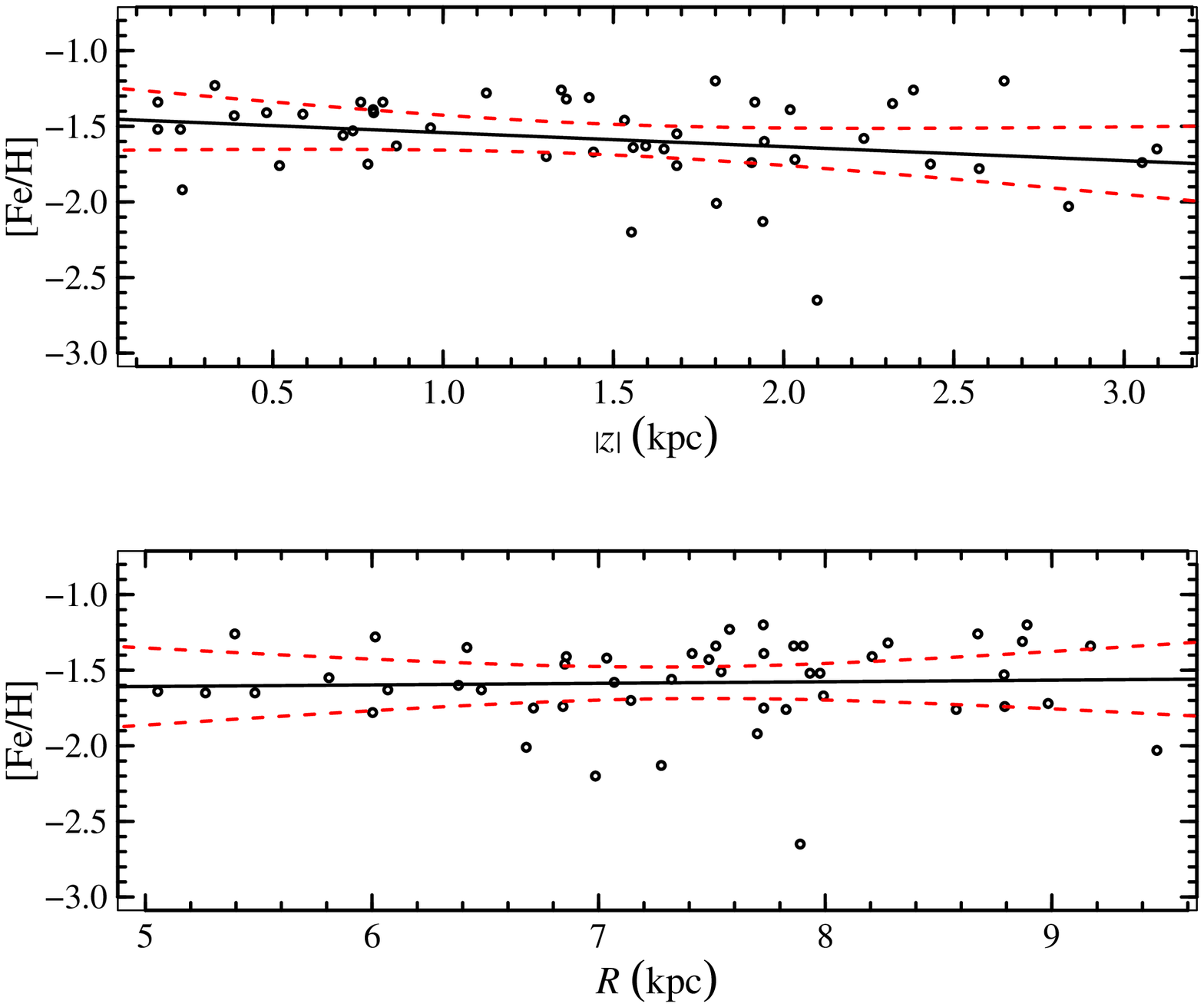}
\caption{$\feh$ vs. $R$ and $|z|$ for stars only assigned to the thick disk (with $\feh<-1.2$) that are not biased by our selection effects, represented as the black points.  Typical errors are $\sim0.1$~kpc in $R$ and $|z|$ and 0.1~dex in the $\afe$ ratios.  The black line is the robust least squares fit to the data, while the red-dashed curves represent the 95-percentile confidence intervals for the fit.  Note that a slope equal to zero (no gradient) is consistent with our data for both $R$ and $|z|$.}
\label{fig-zgfit}
\end{figure}

\begin{figure}[h]
\epsscale{1.1}
\plotone{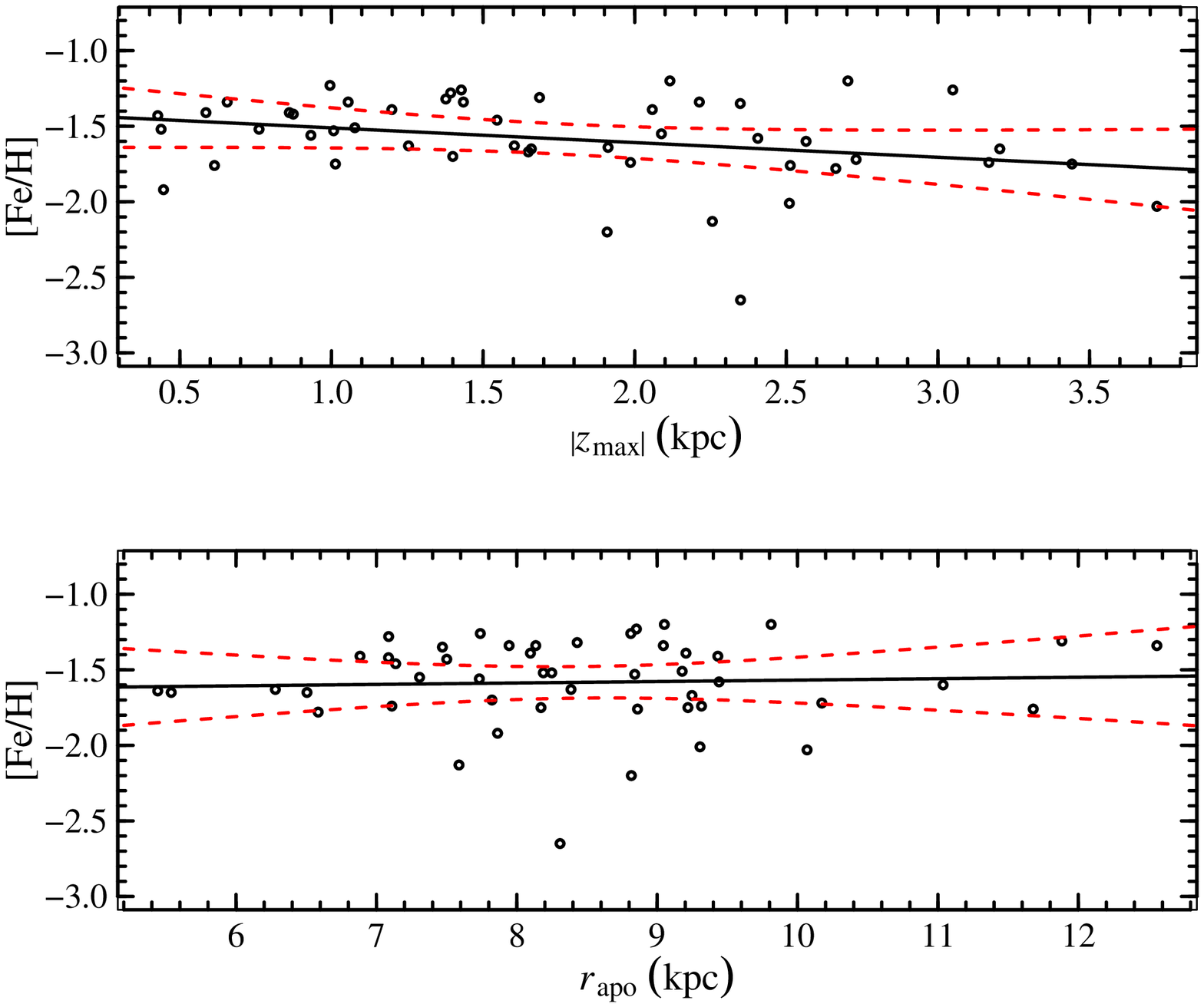}
\caption{$\feh$ vs. $\rapo$ and $|z_{\rm max}|$ (columns 4 and 5 of Table~\ref{tab-orb}) for stars assigned to the thick disk with $\feh<-1.2$.  Typical errors are $\sim0.1$~kpc in $\rapo$ and $|z_{\rm max}|$ and 0.1~dex in the $\afe$ ratios.  The black line is the robust least squares fit to the data, while the red-dashed curves represent the 95-percentile confidence intervals for the fit.}
\label{fig-zgmax}
\end{figure}

Only the 49 stars assigned to the thick disk with $\feh<-1.2$
were used to assess the amplitude of metallicity gradients
in our metal-poor thick-disk sample, as shown in
Figure~\ref{fig-zgfit}.  This figure shows the robust
least squares fits to the data along with 95\% (2~sigma) confidence
intervals for each fit.  Both fits have slopes which are formally non-zero, but not significant, corresponding
to a $-0.09\pm0.05~{\rm dex~kpc}^{-1}$ gradient in the vertical
direction and a $+0.01\pm0.04~{\rm dex~kpc}^{-1}$ gradient in the
radial direction.  Iron abundance versus
$|z_{\rm max}|$ and $\rapo$ (values are maxima for all integrated orbits) also exhibits similar
gradients (see Figure~\ref{fig-zgmax}).

As stated previously, the population assignments are susceptible to possible misassignments.  We therefore checked our results using our PDF values.  The metallicity with the maximum sum of thick-disk PDFs inside a specific velocity bin was chosen as the metallicity in that given bin.  We then fit a line across the maxima to determine possible gradients.  The results showed very similar results are very similar to those found using only stars assigned to the thick disk.

It is important to note that at low iron abundance and large distances we are hindered by
small-number statistics for the thick disk.  We therefore set up a bootstrap analysis, in which we created 10,000 re-samples
consisting of 25 stars randomly selected from the 49 stars assigned to
thick disk with $\feh<-1.2$.  We performed a least squares fit
to each re-sample, and then took the average and standard deviation of
all resamples to determine the range to which the fit is affected by
possible outliers.  

\begin{figure}[h]
\epsscale{1.1}
\plotone{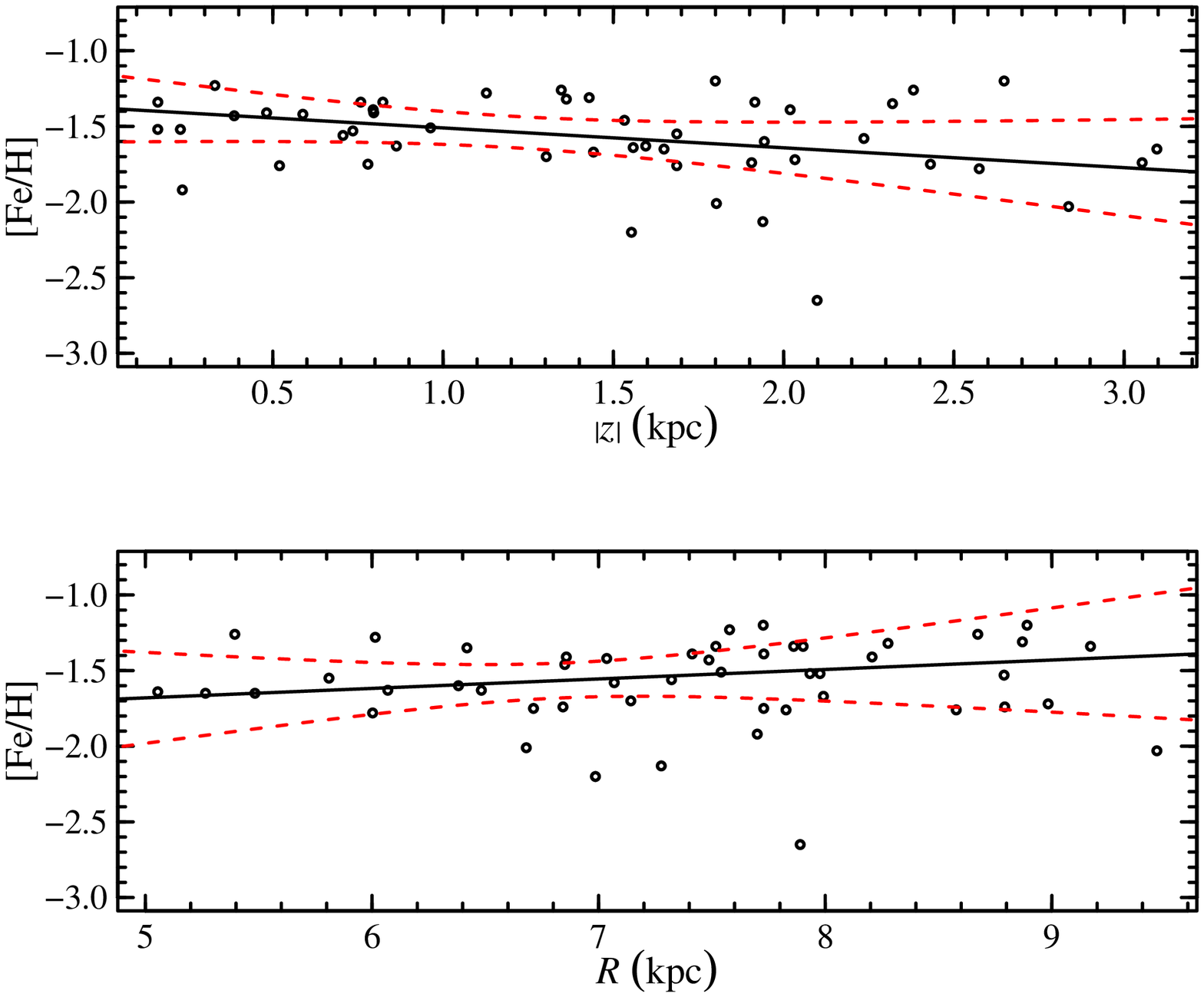}
\caption{$\feh$ vs. $R$ and $|z|$ for stars assigned to the thick disk
(with $\feh<-1.2$).  The black lines are now the mean of least
squares fits to 10,000 random samples, consisting of only half of the
stars assigned to the thick disk.  The red-dashed curves represent the 95\%
confidence intervals derived from the scatter in the distribution of
linear fits to each random sample.}
\label{fig-zgboot}
\end{figure}

Figure~\ref{fig-zgboot} shows the results from this test.  The
confidence intervals shown in this figure represent the degree to
which the fit changes for different random samples.  The mean slopes
are now only slightly steeper than for our original fits, giving a
vertical gradient of $-0.13\pm0.07~{\rm dex~kpc}^{-1}$ and a
$+0.06\pm0.06~{\rm dex~kpc}^{-1}$ radial gradient.  Similarly to our original fits, a slope equal to
zero is also still consistent within errors.

\subsection{Alpha-to-Iron Ratios}

\begin{figure}
\epsscale{1.1}
\plotone{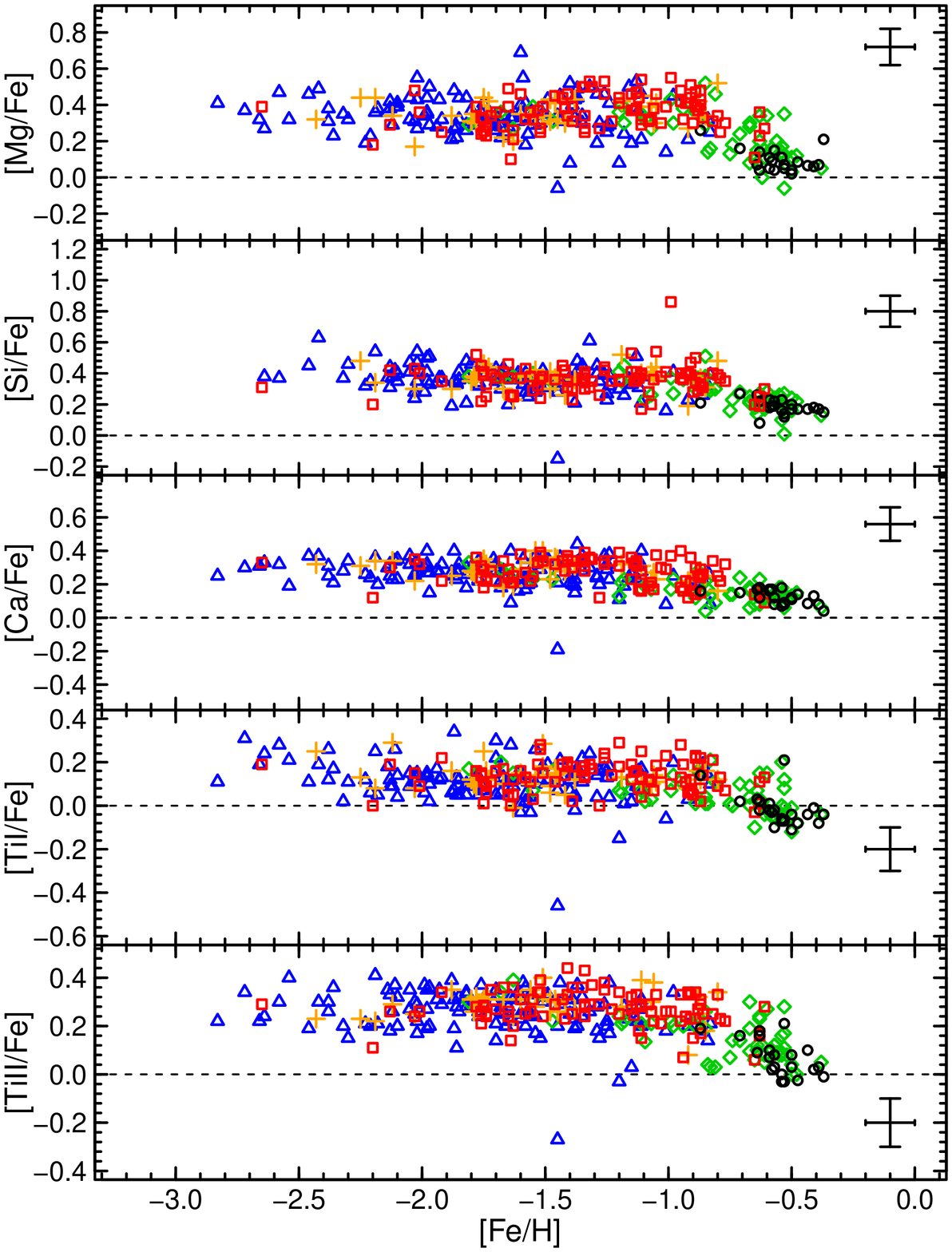}
\caption{Computed $\afe$ ratios vs. $\feh$ for our sample.  Note that element ratios are computed using the iron abundance of the same ionization state as the $\alpha$-element (e.g., [Si/Fe] = [\nion{Si}{I}/\nion{Fe}{I}]) as is suggested by \citet{kraft03}.  Color and symbols are the same as in Fig.~\ref{fig-toom}.  The cause of the offset between [\nion{Ti}{I}/Fe] and [\nion{Ti}{II}/Fe] is most likely non-LTE effects present within the abundance analysis, but the thick disk and halo still show similar enhancement in each.}
\label{fig-alf}
\end{figure}

Figure~\ref{fig-alf} displays several $\afe$ ratios versus $\feh$ for our entire sample, which can be directly compared to the same plots in R10 for only the giant stars.  Most of the stars with $\feh>-0.7$ consist of MS/SG stars that were assigned to either the thin disk or thin/thick
population, and have an $\alpha$-enhancement lower than for the stars at lower metallicities.  The more metal-poor MS/SG stars, however, typically have similar $\alpha$-enhancement to the giants of the same metallicity, an indication that the results from R10 were not affected by the addition of the MS/SG stars.

The metal-poor thick-disk stars have $\afe$ significantly above solar, $\sim0.36$~dex for Mg and Si and $\sim0.27$~dex for Ca and \nion{Ti}{II}.  The mean [\nion{Ti}{I}/Fe] value is about a tenth of a dex lower than that for \nion{Ti}{II}.  This offset is most likely due to non-LTE effects present within our analysis \citep[see][]{bergemann11}.  The ratios also show low scatter, $\simlt0.09$~dex for all $\alpha$-elements, which is less than the 0.1~dex experimental error in $\afe$.  The $\afe$ ratios also blend smoothly into the halo stars with the difference in mean $\afe$ ranging between $0.00-0.03$~dex, well within our experimental error.  There are a few thick-disk stars with $\feh>-0.7$ that may have lower enhancement than the metal-poor thick-disk stars, but they do not represent a large enough sample to make any clear conclusion.  Also note that, as was shown in R10, there is a thick disk star at $\feh\sim-1$ with very large [Si/Fe] enhancement and a halo star at $\feh\sim-1.45$ that shows consistently low $\alpha$-enhancement.  These stars, however, show no peculiarities in their kinematics, and will be the subject of future papers.

\begin{figure}
\epsscale{1.1}
\plotone{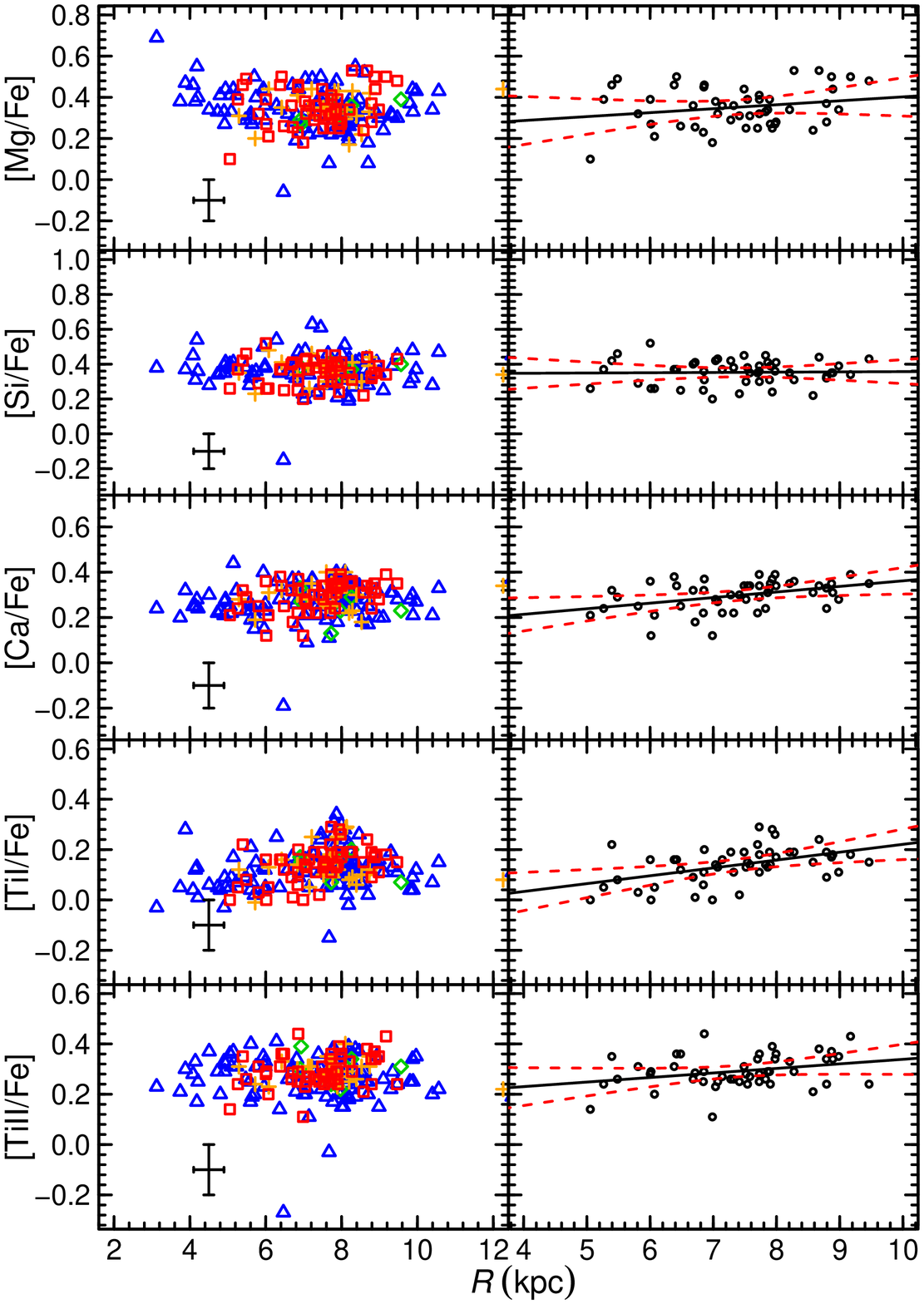}
\caption{Computed $\afe$ ratios vs. $R$ for our sample of
stars with $\feh<-1.2$.  The
left plots show all populations with symbols and colors the same as Figure~\ref{fig-alf}.  The right
plots show the least squares fits to only stars assigned to the thick
disk.  The black line is the fit, while the red-dashed curves
represent the 95\% confidence intervals.}
\label{fig-agr}
\end{figure}

\begin{figure}
\epsscale{1.1}
\plotone{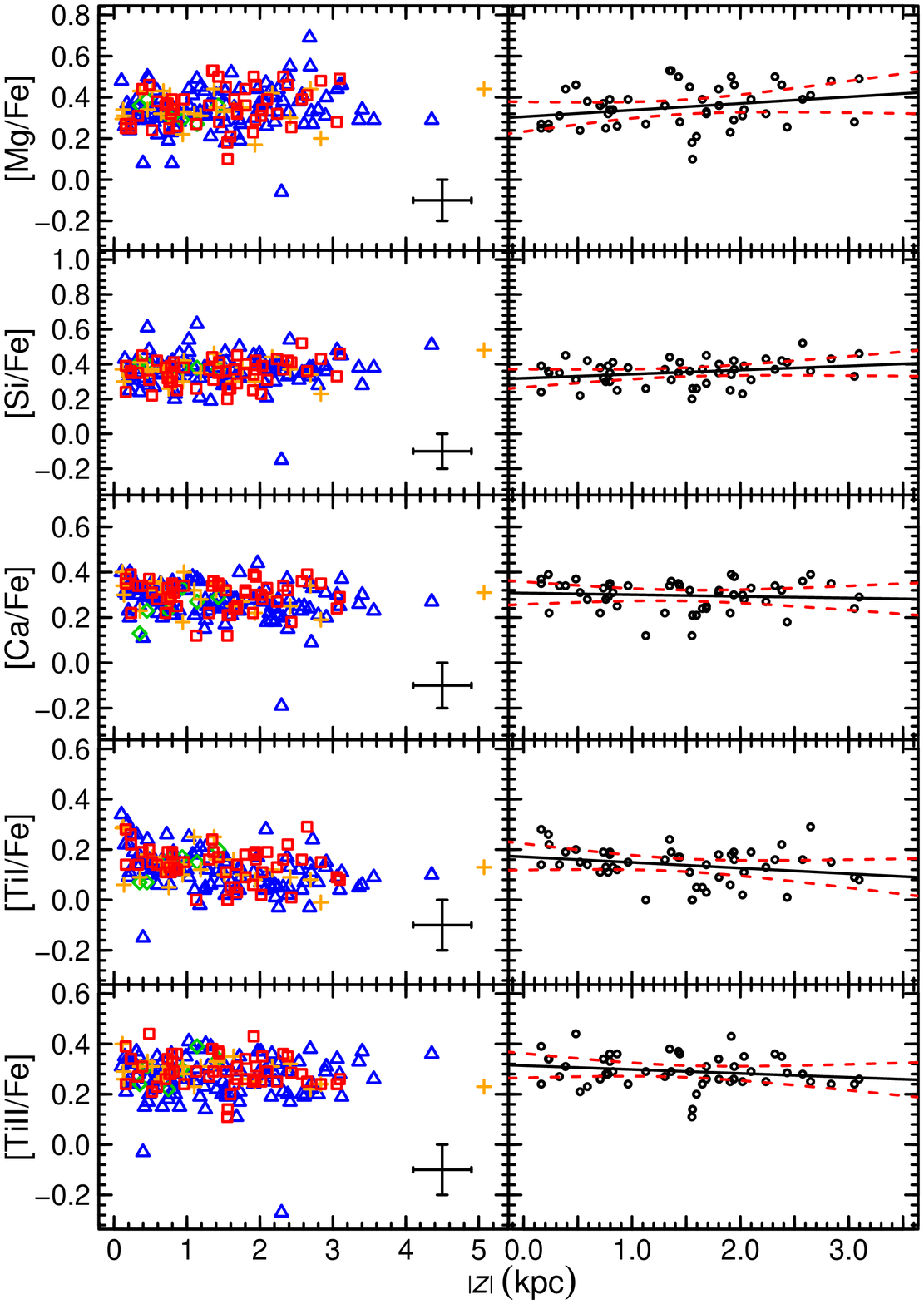}
\caption{Computed $\afe$ ratios vs. $|z|$ for our sample of stars with $\feh<-1.2$.  Symbols and colors are the same as Figure~\ref{fig-agr}.  Recall that the left-hand side shows all populations, while the right-hand side shows only the stars assigned to the thick disk.}
\label{fig-agz}
\end{figure}

\begin{figure}
\epsscale{1.1}
\plotone{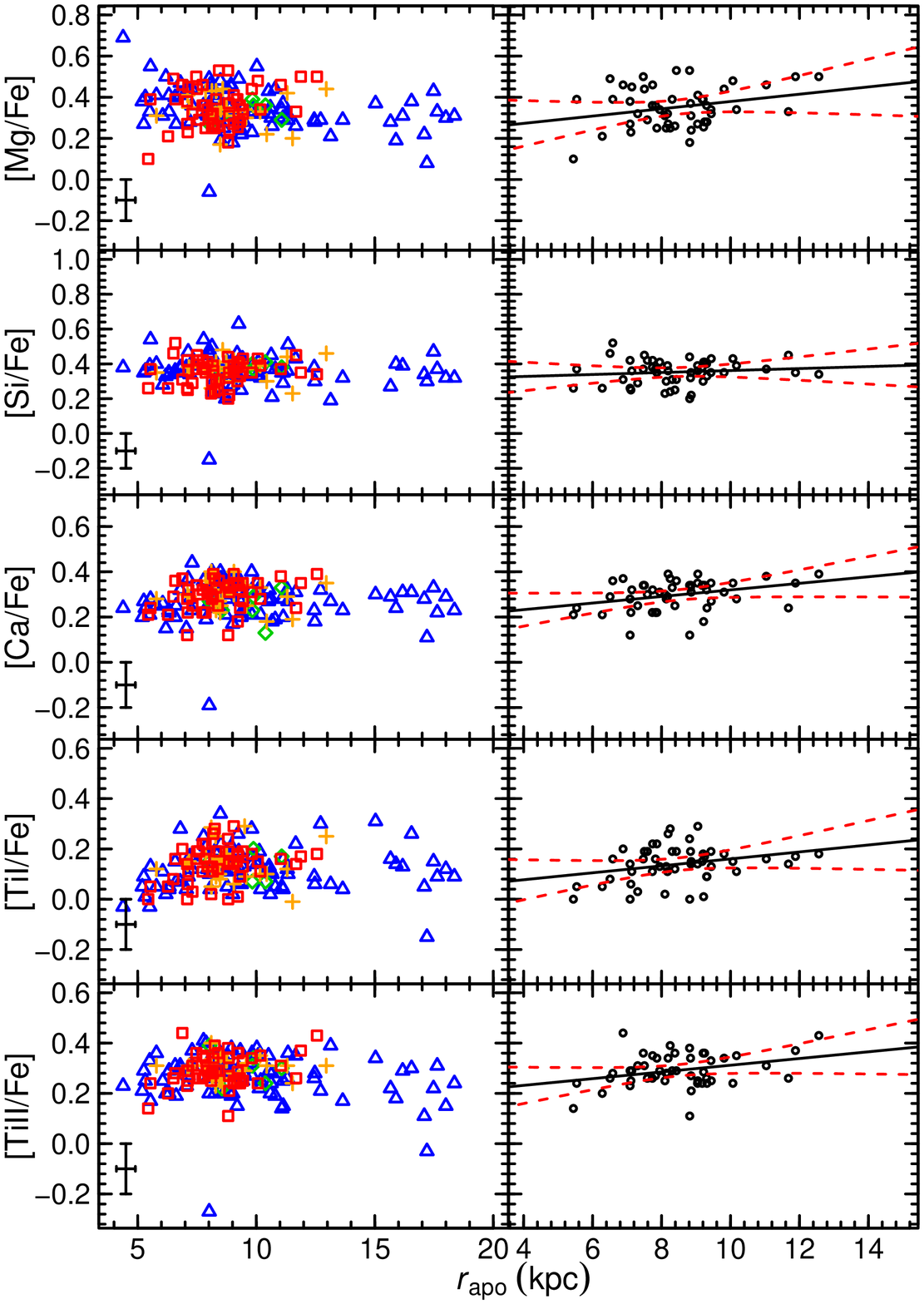}
\caption{Computed $\afe$ ratios vs. $\rapo$ (column 4 of Table~\ref{tab-orb}) for our sample of stars with $\feh<-1.2$.  Symbols and colors are the same as Figure~\ref{fig-agr}.  Recall that the left-hand side shows all populations, while the right-hand side shows only the stars assigned to the thick disk.}
\label{fig-arapo}
\end{figure}

\begin{figure}
\epsscale{1.1}
\plotone{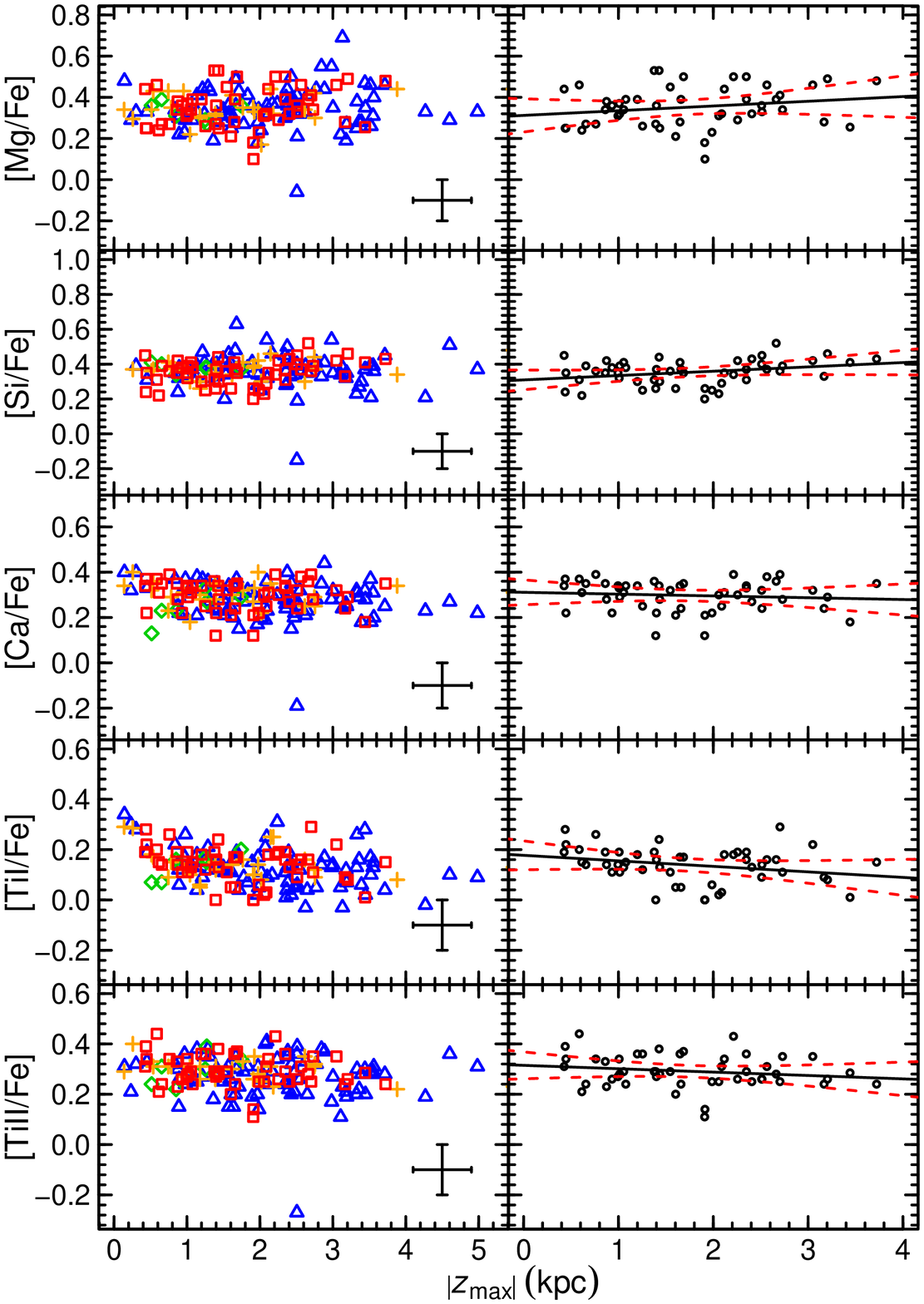}
\caption{Computed $\afe$ ratios vs. $|z_{\rm max}|$ (column 5 of Table~\ref{tab-orb}) for our sample of stars with $\feh<-1.2$.  Symbols and colors are the same as Figure~\ref{fig-agr}.  Recall that the left-hand side shows all populations, while the right-hand side shows only the stars assigned to the thick disk.}
\label{fig-azmax}
\end{figure}

Figures~\ref{fig-agr} and \ref{fig-agz} show the $\afe$ ratios
versus $R$ and $|z|$, respectively, for only stars with $\feh<-1.2$.  In both plots, the stars assigned to the thick disk show little dependence on position,
for all $\alpha$-elements.  In the vertical
direction, [Mg/Fe] and [Si/Fe] slightly increase towards larger $|z|$ at
$+0.03\pm0.02~{\rm dex~kpc}^{-1}$ and $+0.02\pm0.01~{\rm
dex~kpc}^{-1}$, respectively, while [Ca/Fe] decreases at a rate of
$-0.01\pm0.01~{\rm dex~kpc}^{-1}$ and [\nion{Ti}{I}/Fe] and [\nion{Ti}{II}/Fe]
decrease at a rate of $-0.02\pm0.02~{\rm dex~kpc}^{-1}$.  The ratio of all five
elements to iron increase at less than $0.03\pm0.01~{\rm dex~kpc}^{-1}$ radially
outward.  As in the case of iron abundance, the $\afe$ ratios
versus $\rapo$ and $|z_{\rm max}|$ show similar gradients (see
Figures~\ref{fig-arapo} and \ref{fig-azmax}), only shifting by
$\simlt0.01~{\rm dex~kpc}^{-1}$ vertically. In addition, the radial
gradient reduced to $\sim0.01~{\rm dex~kpc}^{-1}$ for all
$\alpha$-elements.

\subsection{High Angular Momentum Halo}
\label{sec-hianghalo}

In Figure~\ref{fig-dvel}, it is evident that some stars assigned to the thick disk have overlapping $V_{\Theta}$ velocities with the high-velocity tail of the halo.   Further, we found that there is a second, low-metallicity peak in the metallicity distribution of the thick disk (Figure~\ref{fig-zdists}).

\begin{figure}
\epsscale{1.1}
\plotone{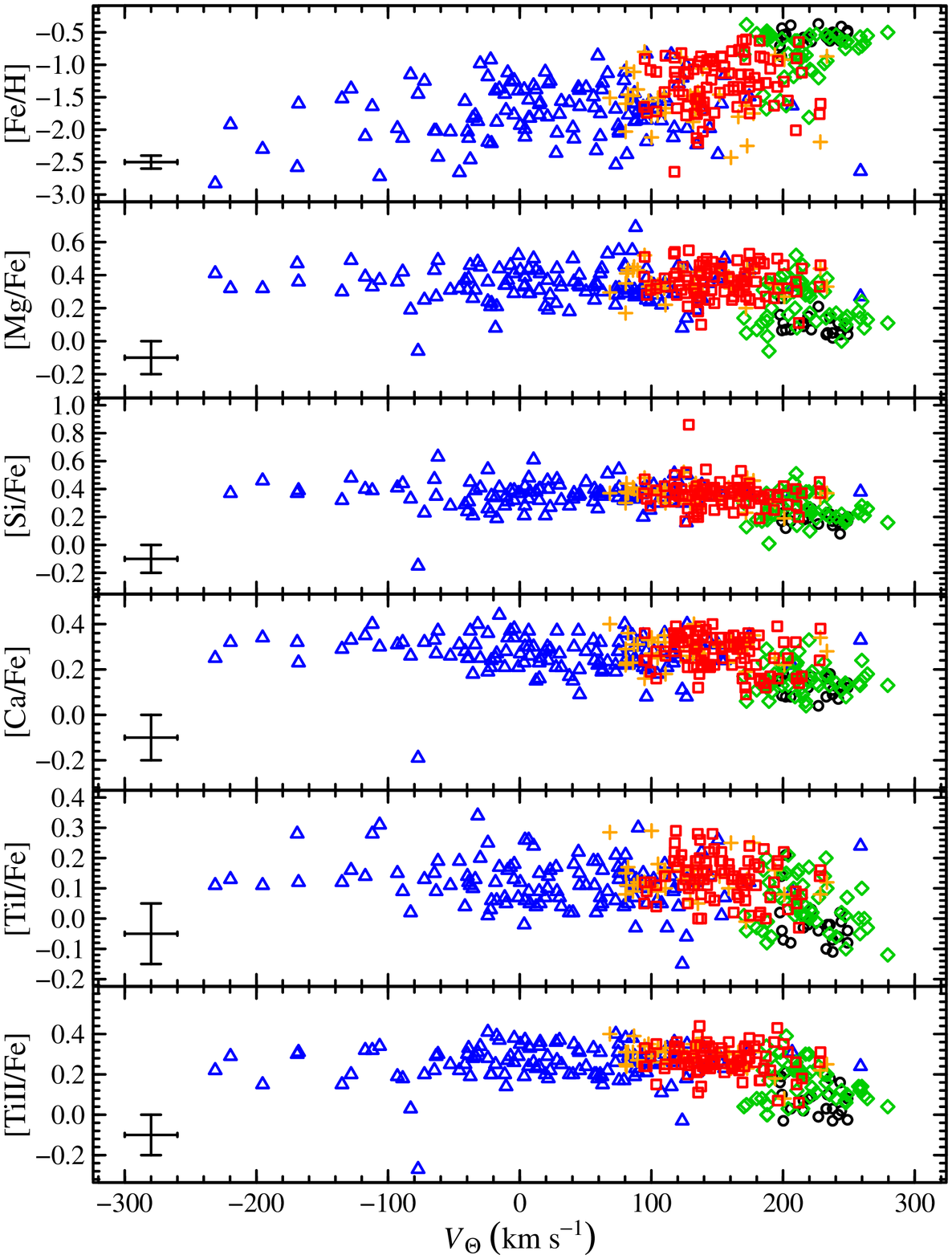}
\caption{$\feh$ and $\afe$ ratios vs. $V_{\Theta}$ for our sample of stars.  Symbols and colors are the same as Figure~\ref{fig-alf}.  Note that the error in $V_{\Theta}$ is typically lower than the error bar shown.}
\label{fig-vgrad}
\end{figure}

We investigated the possibility that the low metallicity peak might indicate contamination from the high-angular momentum halo by plotting $\feh$ and each of the $\afe$ ratios versus $V_{\Theta}$, which are shown in Figure~\ref{fig-vgrad}.  Within the thick disk, there is no trend of $\afe$ or $\feh$ with $V_{\Theta}$, except perhaps at the regime when the thick disk overlaps with the thin disk.  This implies that there is no difference between those stars that might kinematically be a part of the tail of the halo and those that have azimuthal velocities that are too high to be halo.  Thus, Figure~\ref{fig-vgrad} shows there is no difference between the halo and thick disk, as was found when comparing the $\afe$ ratios versus $\feh$.

\begin{figure}
\epsscale{1.1}
\plotone{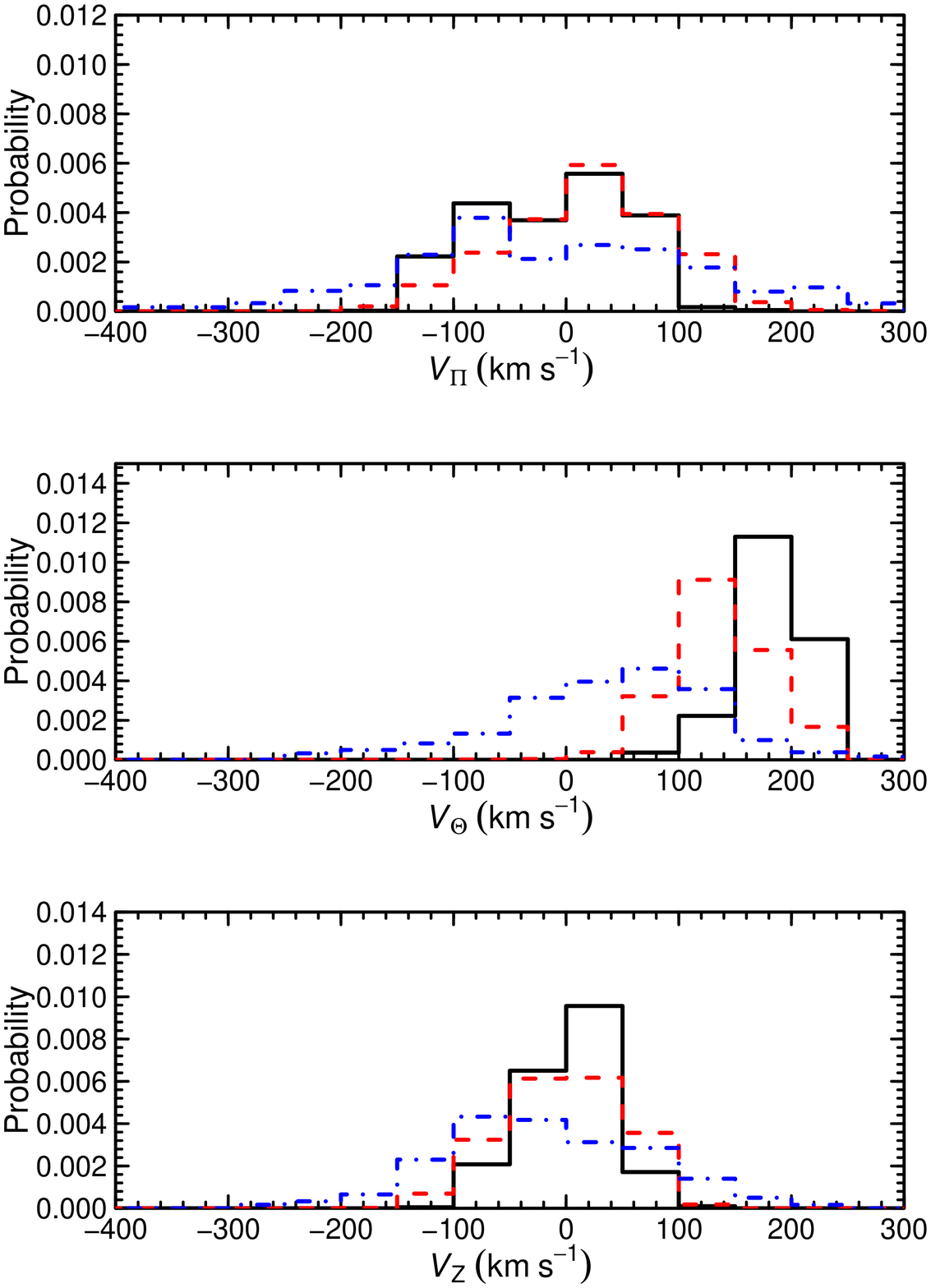}
\caption{Distribution of $V_{\Pi}$ (top panel), $V_{\Theta}$ (middle panel), and $V_Z$ (bottom panel) velocities for stars with $\feh<-1.2$, given the PDF value that they belong to the thin disk (solid black), thick disk (dashed red), and halo (dot-dashed blue).  Note that each distribution was created by summing the PDF values within a given velocity bin.  The distributions were then normalized such that the total area equals unity.}
\label{fig-lowmetdv}
\end{figure}

In Figure~\ref{fig-lowmetdv}, we plot the velocity distributions for only stars with $\feh<-1.2$.  The thick disk distributions still show a clear distinction from the halo distributions.  It is therefore reasonable to assume these stars could still be thick disk stars.

\subsection{A Lagged Thick Disk Component?}

Some have proposed \citep[e.g.,][]{carollo10} that the kinematics of the metal-poor thick disk may differ from the canonical thick disk.  In this case the metal-poor component would have higher velocity dispersions and a slower rotational velocity.  To investigate the effects of such a component on our results, we adopted a $\langle V_{\Theta} \rangle=-100$~\kmsec~and velocity dispersions of $(\sigma_{\Pi}, \sigma_\Theta,
\sigma_Z)=(63, 70, 60)$~\kmsec~for the lagged component \citep[see][]{gilmore02} in addition to the thin disk, canonical thick disk, and halo components.  We then computed new population assignments for the stars.  This increased the number of stars assigned to the thick disk population by 22 stars; 16 of which were formerly assigned to the thick/halo population, 4 formerly halo, and 2 formerly thin/thick.  

The overall shape of the iron distribution did not change for the combined thick disk components.   Further, the mean $\afe$ ratios showed no difference from that of the canonical thick disk alone, while the gradients in both $\feh$ and $\afe$ changed by less than $0.01~{\rm dex~kpc}^{-1}$. Our results therefore do not show any significant change by including a lagged component for the thick disk.  

\section{IMF Variation}

In R10, we established that our metal-poor thick-disk giants
stars formed during a period of rapid star formation, primarily
pre-enriched by core-collapse supernova (e.g. SNe~II).  In this section we quantify the level of IMF
variation (specifically the slope, $x$, of the IMF) of the massive
stars that ended as a core-collapse SNe using model yields and comparing with the
scatter in our data. 

For this test, we adopted the mass-dependent Mg and Fe yields of SNe~II from \citet{kobayashi06}.  we can then compute the massive-star IMF-averaged yield for each element using different IMF slopes, $x$, by:
\begin{equation}
Y_{\rm IMF} = \int_{M_{\rm lo}}^{M_{\rm up}} Y(M)~M^{-(1+x)}~dM
\label{eq-imfint}
\end{equation}
where $Y(M)$ is the mass-dependent yield, $M_{\rm lo}=13$~\msun, and
$M_{\rm up}=40$~\msun.  

\begin{figure}[h]
\epsscale{1.1}
\plotone{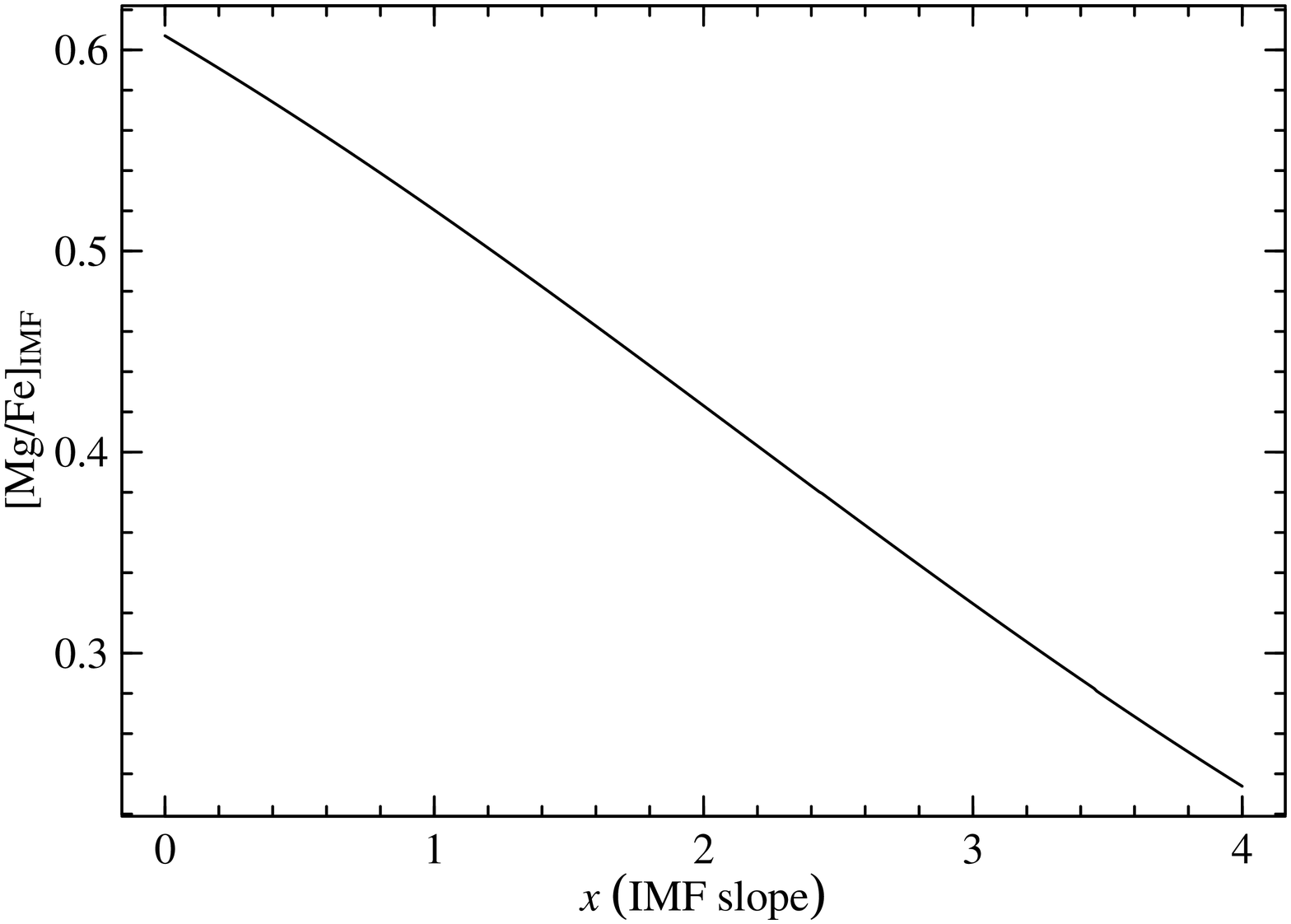}
\caption{IMF-averaged ${\rm [Mg/Fe]_{IMF}}$ vs. the IMF slope, $x$, computed using equation (\ref{eq-imfint}).  Note that as $x$ decreases, the ratio of Mg-to-Fe increases, and that the relationship is nearly linear.  The Salpeter IMF occurs for $x=2.35$.}
\label{fig-imfmod}
\end{figure}

Figure~\ref{fig-imfmod} shows a plot of ${\rm [Mg/Fe]_{IMF}}$ versus the IMF slope,
$x$.  We used this plot to determine the spread in IMF slope
values from the scatter in the [Mg/Fe] ratios of the stars assigned to the thick disk.

From Figure~\ref{fig-imfmod}, a small difference in IMF slope implies
a large difference in $\afe$ (in this case ${\rm [Mg/Fe]_{IMF}}$), better than scatter.  Previously,
we showed that the thick disk has no scatter outside of random errors
in all $\afe$ ratios.  This leaves no room for any variation in the IMF.  Further, this implies that the ISM was well-mixed prior to star formation.
Similarly, the difference between the mean [Mg/Fe] values of the halo
and thick disk is 0.03~dex, well within our 0.1~dex errors.  This
illustrates that the halo and metal-poor thick disk came from a very
similar massive star IMF.

\section{Orbital Eccentricity}
\label{ecc}

\citet[][hereafter S09]{sales09} investigated the utility of orbital
eccentricity ($\varepsilon$) distributions as a tool for
distinguishing several scenarios for the formation of the thick disk.
They compared the predictions from several simulations of the formation of thick disks and found that the $\varepsilon$-distributions also provide a robust
diagnostic to distinguish between stellar populations that form in the simulated galaxy ({\it in situ}) versus those that are formed in a satellite galaxy and accreted into the simulated galaxy for all scenarios that involve both types of populations.
The accreted population dominates high eccentricities, while the {\it
in situ} population dominates the lower eccentricity bins.  It is
important to note that only one realization, with a specific
set of initial conditions, for each formation scenario was used to
compute the eccentricity distributions.  This is, however, sufficient since there is no reason that the given simulations are not representative of that scenario.

It is important to note that we cannot use our population assignments when constructing the distribution of orbital eccentricities for the thick disk.  The shape and extent of the tail of the thick-disk distribution is biased by our {\it definition} of `thick disk' during our population analysis.  We simulated our analysis by creating 1000 model stars in which their distance and Galactic coordinates were randomly chosen assuming a uniform distribution, and their 3-dimensional space motion was randomly selected using the `thick disk' Gaussian definition in Table~\ref{tab-gaus}.  Each model star was then run through our orbital program to determine the orbital eccentricity.  Only 12\% of these model thick disk stars had $\varepsilon>0.6$, which indicates that our definition of thick disk does not allow for many stars on highly eccentric orbits.  Further, this could also affect the shape of the right side of the eccentricity distributions.  

\begin{figure}[h]
\epsscale{1.1}
\plotone{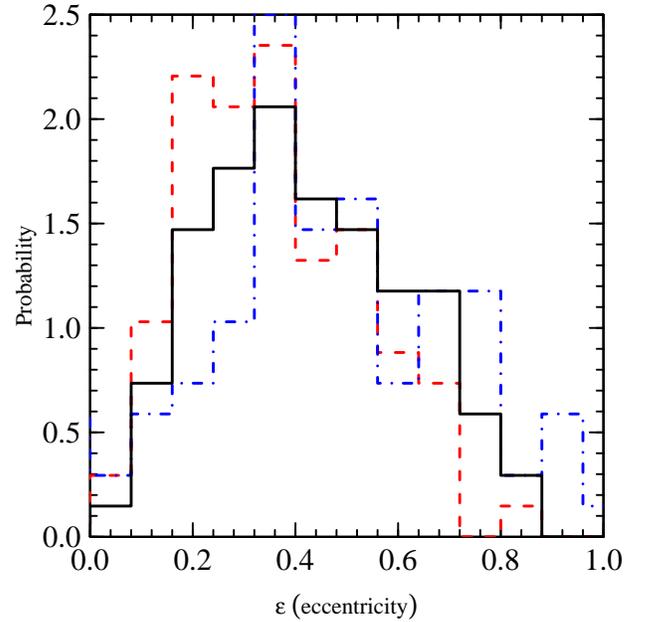}
\caption{The distribution of orbital eccentricities for those stars in our sample with $V_{\Theta}>50$~\kmsec and $1\leq|z|/z_{\rm d}\leq3$, where $z_{\rm d}=0.9$~kpc.  The distribution when the distances to the stars are decreased or increased by 20\% are given by red, dashed and blue, dot-dashed lines, respectively.}
\label{fig-esales}
\end{figure}

We therefore followed the procedure of S09 to investigate the orbital eccentricities of our metal-poor thick disk sample stars, calculated in \S\ref{velorb} for the last orbit of a star.  In this case, we computed the $\varepsilon$-distribution of stars with $V_{\Theta}>50$~\kmsec and $1\leq|z|/z_{\rm d}\leq3$, where $z_{\rm d}=0.9$~kpc.  This distribution is shown in Figure~\ref{fig-esales}.  To estimate the effects of distance errors on the distribution, we also plot the resultant eccentricity distribution when distances are increased and decreased by 20\%.

The majority of the stars in this sample appear to exhibit low orbital eccentricities, with the distribution peaking around $\varepsilon\sim0.3-0.4$.  Our distribution does not show a strong resemblance to any of the distributions found in S09.  The tail to high eccentricity somewhat resembles that of the direct accretion scenario of \citet{abadi03}.  The peak at lower eccentricities is significantly lower than predicted by the direct accretion scenario, and is more consistent with the predictions of models wherein the stars of the thick disk were primarily formed {\it in situ}.  It is possible that we are under-sampling the low-eccentricity bins since we do not have many stars at the peak of the thick-disk metallicity distribution, $\feh\sim-0.6$ .  It is likely that including more metal-rich thick-disk stars could significantly change the distribution of eccentricities.  

\section{Discussion}

In this work, we analyzed our full sample of metal-poor thick disk candidate stars.  The MS/SG stars, added to our giant star sample from R10,
were primarily assigned to the thin disk and thin/thick populations, and increased the number of stars with $\feh>-0.8$.  We found that our findings from R10 were unchanged, in that the $\afe$ ratios
for the metal-poor thick-disk stars are enhanced and show low scatter
($<0.09$~dex, within the error of 0.1~dex), indicating that star
formation took place on a short timescale in which the metal-poor
thick disk was pre-enriched by core-collapse SNe from an invariant
massive star IMF.  Further, the metal-poor thick disk and halo were
most likely pre-enriched by the same massive star IMF, showing a
difference in $\alpha$-enhancement of $<0.03$~dex.  The low amplitude of scatter in the element ratios indicates that the ISM from which the stars formed was well-mixed.

As discussed in R10, the enhancement and low scatter of
$\afe$ in the thick disk are evidence that the formation of the
thick disk had little influence by the late, direct accretion of stars from
dwarf galaxies.  The $\alpha$-enhancement in the metal-poor thick disk
contrasts with the expectations from models that have
direct accretion up until about $\sim6$~Gyr ago
\citep[e.g.,][]{abadi03}, assuming that the accreted
dwarfs formed stars and self-enriched similarly to the surviving
dwarfs.  In this case, the stars accreted would then look
chemically different from our metal-poor thick-disk stars (see
R10).  Further, the distribution of orbital eccentricities for our stars does not resemble that of the direct
accretion scenario, instead resembling a population that was formed {\it in situ}.  

Using the full RAVE catalog of stars, \citet{wilson11}
also computed orbital eccentricities for a more uniform sample of thick disk stars,
finding results consistent with ours.  On the other hand, \citet{dierickx10} looked at the orbital eccentricities of a sample of thick-disk candidate stars selected from the Sloan Digital Sky Survey, and found that their distribution is inconsistent with the thick disk forming from radial migration only or heating due to mergers only.  \citet{casetti-dinescu11}, however, show that their distribution, using data from RAVE and the Southern Proper Motion Program, is most consistent with the heating and merger scenarios.  Overall, these results still support our finding that the stars in the thick disk primarily formed {\it in situ}.  It is also important to note that our distribution of eccentricities is the first to have the metal-poor thick disk well-represented, but selection biases may be affecting the shape of the distribution.

From previous investigations of the local neighborhood and our more extended sample, it is clear that the thick disk is old, and had to form stars during a short, rapid burst to produce the $\alpha$-enhancement we detect.  Direct accretion of stars from dwarf galaxies that formed a long time ago (about $10$~Gyr ago) during a period of rapid star formation is then still viable.  These types
of dwarf galaxies, however, are extremely rare.  All known dwarf
spheroidal galaxies typically have extended star formation and lower $\alpha$-enhancement at the
same $\feh$ range of the metal-poor thick disk \citep[see review
by][]{tolstoy09}.  We can therefore conclude with confidence that the
{\it late} accretion of stars from satellite galaxies did not play a
role in the formation of the thick disk.

Models that include a significant component of the thick disk formed
{\it in situ}, however, need to be further assessed.  These models all
imply or directly predict a thick disk with high $\alpha$-enhancement.  Further, given the errors in the distances to our stars, and the effects of our definition of `thick disk', the orbital eccentricity distribution of our sample of metal-poor thick disk stars does not show any direct inconsistencies with the distributions in S09 for these scenarios.  It is interesting to note, however, that the S09 $\varepsilon$-distribution for the heating scenario shows a significant fraction of accreted stars that comprise the thick disk.  Unless all of these accreted stars were assimilated into the thick disk at early times (as stated above), then this fraction of stars would have lower $\alpha$-enhancement than we see in our sample.  

The models also show
differences in their predictions for radial or vertical abundance
variation.  Observational evidence for abundance gradients in the
thick disk are therefore very important.  The majority of previous studies with high resolution analyses have
shown no evidence for a vertical metallicity gradient in the thick
disk \citep[cf.,][]{mishenina04,soubiran05}.  These studies, however,
primarily include nearby stars with $\feh\geq-1.5$, while our
gradient is computed for thick-disk stars with $\feh<-1.2$.
More recent studies have shown evidence for the possibility of a
vertical metallicity gradient.  Looking at stars at high Galactic
latitudes and $z<4$~kpc, \citet{siegel09} found that iron abundance
decreases with vertical height by $-0.15~{\rm dex~kpc^{-1}}$.  This
data, however, may have strong contamination from the thin disk.  \citet{ivezic08} found a $0.1-0.2~{\rm dex~kpc^{-1}}$ metallicity gradient for stars selected from SDSS that lie a vertical heights, $z=1-2$~kpc.  At these heights, the likelihood of contamination from thin disk is small, and so one could deduce that they were looking at mostly thick disk stars.

The stars in our sample assigned to the thick disk lie within about
$\pm2$~kpc radially about the Sun, and have vertical distances of
$|z|<3$~kpc.  Due to biases introduced by our analysis technique, we only investigated abundance
gradients for those stars with $\feh\leq-1.2$.  We found very
small amplitude radial and vertical gradients, $<0.03\pm0.02~{\rm
dex~kpc^{-1}}$, in the $\alpha$-enhancement of the metal-poor thick
disk. This further verifies that the ISM was well mixed during star
formation.  Further, we found a small $+0.01\pm0.04~{\rm dex~kpc^{-1}}$
radial iron abundance gradient, however, it is possible that $\feh$
significantly changes by $\sim-0.09\pm0.05~{\rm dex~kpc^{-1}}$ with vertical
height above the Galactic plane.  A fit resulting in no gradient is
still possible, however, within 2-sigma confidence limits.  Note that it is possible that including the more metal-rich thick disk would significantly change the amplitude of the slope of the gradients.  Our results, however, resemble those found by groups studying the more metal-rich thick disk (as given above).

A vertical metallicity gradient is expected from the dissipational
collapse model.  Additionally, a thick disk with uniformly enhanced $\alpha$-abundances is also
likely if the collapse took several millions of years, as predicted \citep{burkert92}.  Rapid star formation early on during the heating scenario will produce
enhanced $\afe$ ratios, while small amplitude radial abundance
gradients are also possible.  The merger model also
predicts a high star formation rate from gas dissipation, and hence
enhanced $\afe$ ratios, but the final thick disk is expected to
have uniform abundance ratios.  Although our data does not directly
disagree with a uniform disk, a significant iron abundance gradient
would challenge this scenario.  It is also important to note that
there is direct accretion of stars during the merger and heating
scenarios, which can occur until late times.  Late accretion of stars
must therefore be extremely minimal for the abundances of the final
thick disks in these scenarios to match the low scatter in
$\afe$ for our thick disk stars.

An $\alpha$-enhanced thick disk is also implied from the clumpy disk
model in which there is rapid star formation from internal processes
with rapid mixing by strong turbulence.  Additionally, the thick disk
in the radial migration model has high $\afe$ ratios, since the stars in the inner disk undergo rapid star formation and then move outward, \citep[see][]{schonrich09b}.  No radial abundance gradients are predicted due to blurring across radii
in the radial migration model and the turbulent mixing in the clumpy
disk model.  Note that the radial migration model does not include effects from the bar of the Galaxy.  It is possible that the combination of the bar and spiral arms could result in a very efficient mixing mechanism, which could cause variation in the metallicity of the thick disk as a function of radius \citep{minchev10,minchev11}.  It is unclear, however, if a vertical iron abundance
gradient is possible in either the radial migration model or clumpy disk model, indicating that more modeling
must be done to ascertain this chemical signature.

\section{Conclusions}

The metal-poor thick disk of the Milky Way Galaxy is enhanced in the
$\alpha$-elements and reaches to metallicities down to -2~dex.  We find that the stars in the thick disk most likely formed within the potential
well of the Milky Way Galaxy.  Direct accretion of stars could have
occurred at very early times ($\sim1$~Gyr after the start of star
formation) in the formation of the thick disk, but the later
contribution of accreted stars into the thick disk was very minimal.
The abundance trends of the metal-poor thick disk tends to favor
models which result in a thick disk with a significant vertical
metallicity gradient, however, a uniformly enhanced thick disk is
still possible.  Additionally, the radial migration and internal star
formation at high redshift may have also contributed to the formation
of the thick disk, but more work needs to be done to quantify the
abundance trends for these scenarios.

\acknowledgements
 We would like to thank the staff members of Siding Spring Observatory, 
Apache Point Observatory, ESO La Silla Observatory, and Las Campanas Observatory for their assistance 
in making these observations possible.  GRR would like to thank K. Lind and M. Bergemann for helpful discussions on non-LTE effects in stellar atmospheres.  GRR acknowledges support from the National Science Foundation of the USA (AST-0908326).  JPF acknowledges support 
through grants from the W. M. Keck Foundation and the Gordon and Betty Moore
Foundation, to establish a program of data-intensive science at the Johns
Hopkins University.  This publication makes use of data products of the
2MASS survey, which is a joint project of the University of
Massachusetts and IPAC/Caltech, funded by NASA and the NSF.  This research
has also made use of the Vizie-R databases, operated at CDS, Strasbourg, France.
Funding for RAVE has been provided by: the Australian Astronomical
Observatory; the Leibniz-Institut f\"ur Astrophysik Potsdam (AIP); the Australian National
University; the Australian Research Council; the French National Research
Agency; the German Research Foundation; the European Research Council 
(ERC-StG 240271 Galactica); the Istituto Nazionale di
Astrofisica at Padova; The Johns Hopkins University; the National Science
Foundation of the USA (AST-0908326); the W. M. Keck foundation; the
Macquarie University; the Netherlands Research School for Astronomy; the
Natural Sciences and Engineering Research Council of Canada; the Slovenian
Research Agency; the Swiss National Science Foundation; the Science \&
Technology Facilities Council of the UK; Opticon; Strasbourg Observatory;
and the Universities of Groningen, Heidelberg and Sydney. The RAVE web site is at http://www.rave-survey.org.

{\it
\facility
Facilities: \facility{ARC (echelle spectrograph)}, \facility{AAT (UCLES)}, \facility{Magellan:Clay (MIKE)}, \facility{Max Planck:2.2m (FEROS)}, \facility{UKST (6dF spectrograph)}
}

\clearpage

\end{document}